\documentclass[11pt]{article}
\usepackage{jcappub}
\usepackage{graphicx}

\usepackage[dvipsnames]{xcolor}
\usepackage{xspace}

\usepackage[normalem]{ulem}

\newcommand{\overbar}[1]{\mkern 1.5mu\overline{\mkern-3.5mu#1\mkern-1.5mu}\mkern 1.5mu}
\newcommand{\diff}{\mathop{}\!\mathrm{d}}
\newcommand{\td}[2]{\frac{\diff #1}{\diff #2}}

\newcommand{\ib}{\mathrm{I}\!B}
\newcommand{\ic}{\mathrm{IC}}
\newcommand{\fnl}{f_\mathrm{NL}}
\newcommand{\fnlloc}{f_\mathrm{NL}^\mathrm{local}}
\newcommand{\fsky}{f_\mathrm{sky}}
\newcommand{\lmax}{\ell_\mathrm{max}}
\newcommand{\nside}{N_\mathrm{side}}

\newcommand{\planck}{\textit{Planck}}

\newcommand{\Thomson}{Thomson\xspace}

\title{The integrated angular bispectrum}
\author[a,b]{Gabriel  Jung,}
\author[a,b]{Filippo Oppizzi,}
\author[c]{Andrea Ravenni,}
\author[a,b]{and Michele Liguori}

\affiliation[a]{Dipartimento di Fisica e Astronomia "G. Galilei", Universit{\`a} degli Studi di Padova,\\ Via Marzolo 8, 35131 Padova, Italy}
\affiliation[b]{INFN, Sezione di Padova, via Marzolo 8, I-35131, Padova, Italy}
\affiliation[c]{Jodrell Bank Centre for Astrophysics, School of Physics and Astronomy, The University of Manchester, Manchester M13 9PL, U.K.}

\emailAdd{gabriel.jung@pd.infn.it}
\emailAdd{michele.liguori@pd.infn.it}

\emailAdd{andrea.ravenni@manchester.ac.uk}


\arxivnumber{}

\abstract{We study the position-dependent power spectrum and the integrated bispectrum statistic for 2D cosmological fields on the sphere (integrated angular bispectrum). First, we derive a useful, $m$-independent, formula for the full-sky integrated angular bispectrum, based on the construction of azimuthally symmetric patches. We then implement a pipeline for integrated angular bispectrum estimation, including a mean-field correction to account for spurious isotropy-breaking effects in realistic conditions (e.g., inhomogenous noise, sky masking). Finally, we show examples of applications of this estimator to CMB analysis, both using simulations and actual \planck\ data. Such examples include $\fnl$ estimation, analyses of NG from secondary anisotropies (ISW-lensing and ISW-tSZ-tSZ bispectra) and studies of NG signatures from foreground contamination.}

\begin{document}

\maketitle
\flushbottom

\section{Introduction}
The bispectrum (three-point function in Fourier space) is the lowest order statistic that carries information about non-linearities in the cosmological density perturbation field, either of primordial or gravitational origin. If we are in presence of weak non-Gaussianity (NG), as is the case when studying primordial non-linearities from Inflation or galaxy clustering on large scales, the bispectrum is also generally the dominant term in a higher order cumulant expansion.

For these reasons, the bispectrum is a powerful cosmological observable. Yet, its estimation using state-of-the-art and future big cosmological datasets is a technically very difficult task. 
The main challenge is posed by the fact that the full bispectrum domain 
is composed by all the closed triangles that can be built using the modes of the cosmological survey under study, e.g. Cosmic Microwave Background (CMB) multipoles or Fourier modes of the galaxy density field. The number of configurations to evaluate is therefore very large for current high resolution observations, and roughly scales like $N_\mathrm{modes}^3$, where $N_\mathrm{modes}$ is the number of modes in the survey (e.g., maximum multipole $\lmax$ or largest wavenumber $k_\mathrm{max}$).

Different methodologies have been developed to deal with this issue. One of the most common approaches in the analysis of CMB data is that of fitting specific bispectrum templates to the data, rather than estimating the three-point function triangle by triangle. If these templates present suitable mathematical properties, namely factorizability, it is possible to build Fast Fourier Transform (FFT)-based bispectrum-fitting algorithms \cite{Komatsu:2003iq,Fergusson:2009nv}. This methodology is particularly useful in tests of primordial non-Gaussianity, where the observed bispectrum has to be compared with theoretical predictions from Inflation. 

Other methods, adopted both in CMB and Large Scale Structure (LSS) analysis, are based on compressing the information by binning the triangles, often followed again by speeding up the estimation via FFT \cite{Scoccimarro:2015bla,Watkinson:2018efd,Bucher:2015ura}. 

The development of these solutions allowed solving the problem of efficient, optimal primordial bispectrum estimation in Cosmic Microwave Background data.
Significant challenges are however still present for Large Scale Structure bispectra. The crucial issue in this case comes from intrinsic non-linearities, arising from gravitational evolution, or from the physics of galaxy formation. Such sources of non-Gaussianity produce a very complex bispectrum covariance structure. For this reason -- along with continued attempts to produce more and more efficient and accurate methods for full LSS bispectrum reconstruction -- alternative observables to test non-linear evolution of structures have been proposed. Such observables typically carry (compressed) information from just a subset of the total amount of available triangle configurations. Therefore, they do not consider the full bispectrum domain. On the other hand, they are built in such a way as to be much simpler and faster to estimate and analyze. 
An important example of this type of statistics, and the one we will be focusing on in this work, is the so-called {\em integrated bispectrum}, originally introduced in \cite{Chiang:2014oga}. The integrated bispectrum can be derived as follows. First, divide the sky in patches and measure the small scale power spectrum in each patch. Then, correlate the {\em position-dependent power spectrum} so obtained with the mean density perturbation in each patch and integrate over the whole sky. In this way, we are building an integrated statistic, which measures how small scale perturbations are modulated by large scale modes. The response of small scale power to large scale fluctuations depend on the {\em squeezed} configurations of the bispectrum, i.e., on bispectrum triangles for which one wavenumber is much smaller than the other two. This method is thus useful to study the squeezed limit of the bispectrum.

So far, the integrated bispectrum has been considered mostly in galaxy clustering and 21cm studies, in the three-dimension (3D) Fourier domain (e.g.\ \cite{Chiang:2015eza,Chiang:2014oga,Chiang:2015pwa,Giri:2018dln}). The integrated bispectrum in two-dimensions (2D) is also very interesting for a variety of applications, such as weak lensing and CMB analyses. The 2D case was considered in \cite{Munshi:2019jyl, Munshi:2016xfg}, where a flat-sky approach was generally taken.

In this work instead, we study in detail the full sky, 2D integrated bispectrum in the multipole domain. 

In the first part of our paper, we introduce two new, useful technical developments. The first one is the construction and use of azimuthally symmetric patches in the analysis. This allows us to derive tractable analytical expressions for the integrated bispectrum, which are very useful for theoretical predictions, and to validate results from actual datasets and simulations. The second development is the explicit introduction of a mean-field correction term in the integrated bispectrum estimator, with aim of removing the increase in error bars, which is always generated in presence of isotropy breaking effects, such as partial sky coverage or spatially inhomogeneous noise. We build our linear mean-field correction following the same approach as for primordial bispectrum or CMB lensing estimators, adapted to the integrated bispectrum case.

In the second part of the paper, we show several concrete applications of the estimator, focusing on CMB data analysis. First, we compute  theoretical integrated angular bispectrum templates for the primordial local and for the ISW-lensing bispectrum shape respectively, obtained by using our azimuthally-symmetric patches. We then measure the integrated angular bispectrum of \planck\ data and fit our theoretical templates to it. In this way, we measure bispectrum amplitude parameters for these shapes and compare our results with those obtained by the \planck\ team, using optimal bispectrum estimators \cite{Ade:2013ydc, Ade:2015ava, Akrami:2019izv}. The reason why we focus on these two bispectra is of course that they peak on the squeezed limit, to which the position-dependent power spectrum is sensitive. Of course the integrated bispectrum estimator --- used in this way for template-fitting --- produces suboptimal results, compared to the full bispectrum analysis (although not by much, especially for the local case, as we will show). However we still believe our analysis of these shapes is useful, both because it allows to check our pipeline against robustly validated \planck\ results, and because the integrated bispectrum allows for a further nice confirmation of such results in the squeezed limit, due to its very direct and transparent physical interpretation (i.e., a direct measurement of the spatial modulation of power in patches). 

Following this analysis, we extensively study another NG signature produced by secondary sources, namely the bispectrum arising from the coupling between the ISW and the thermal Sunyaev Zeldovich (tSZ) effect. 
This shape was recently considered in \cite{Hill:2018ypf}, where the author argues that it can be a potential source of bias for primordial squeezed bispectra, especially for future surveys; of course, this NG signal is also interesting per se, being a specific prediction of the standard cosmological model. Moreover, the ISW-tSZ-tSZ bispectrum peaks in the squeezed limit --- like all three-point signals arising from the coupling of ISW and small scale secondary CMB anisotropies ---  hence it is a good candidate for an integrated bispectrum analysis.
We produce theoretical predictions of the expected integrated bispectrum arising from this shape and its level or primordial NG contamination for future experiments, if not properly removed from the CMB maps. Finally we fit the integrated bispectrum ISW-tSZ-tSZ template to \planck\ data. 

For our last application, we consider diffuse astrophysical foregrounds. It has been argued in recent literature \cite{Jung:2018rgf, Coulton:2019bnz} that the bispectrum could be a useful diagnostic tool of residual foreground contamination in present and future CMB temperature and polarization data. Considering that foreground contamination usually produces a large signature on squeezed triangles, and that one would like a diagnostic tool of this type to be easy and fast to compute, the integrated bispectrum seems particularly well suited to this purpose. As an example we extract the integrated bispectrum from \planck\ maps of galactic thermal dust emission obtained using \texttt{Commander} \cite{Eriksen:2007mx} and discuss its potential applications.

The plan of this paper is as follows. In section \ref{sec:integrated-bispectrum} we start by providing some general definitions and by deriving the general expression of the integrated bispectrum in multipole space. We then discuss our implementation of azimuthally symmetric 2D patches, leading to simple, $m$-independent, analytical formulae for the integrated bispectrum and its covariance. In section \ref{sec:ib-estimator} we describe in detail our integrated bispectrum estimator and its implementation, including the derivation of the mean-field (linear) correction term. In section \ref{sec:data} we discuss the applications of our pipeline to CMB data. Finally, we draw our conclusions in section \ref{sec:conclusion}.

\section{Integrated bispectrum on the sphere}
\label{sec:integrated-bispectrum}

Chiang et al.\ \cite{Chiang:2014oga,Chiang:2015pwa} introduced the position-dependent power spectrum and the so-called integrated bispectrum as a simple way to extract information about the squeezed limit of the bispectrum, by considering only two-point statistics. This method is based on three relatively simple steps:
\begin{itemize}
    \item Separate the sky into several patches of the same size.
    \item In each patch, compute the mean value and the power spectrum of the observed field.
    \item Determine the patch-by-patch correlation of these two quantities: for each patch multiply mean and power and take the average of this product over all the patches.
\end{itemize} 
By construction, this procedure gives a direct access to the modulations of the small-scale fluctuations (power spectrum in a patch) by the large-scale mode (average of a patch), which indeed corresponds to the squeezed limit of the bispectrum.

This section is devoted to the analytical description of the position-dependent power spectrum. First, we recall its formulation on the celestial sphere, as introduced in \cite{Munshi:2019jyl}. Here we adopt a different notation, more usual in the context of CMB analysis. In the second part, we give exact expressions of the integrated bispectrum and its covariance matrix, which are easy to evaluate numerically (as no integral is left). These constitute the main analytical results of this paper and will be used extensively in the following sections.

\subsection{Definitions}
\label{sec:definitions}

Our starting point is a 2D map $M(\hat{\Omega})$ in pixel space for a generic field defined on the sphere, related to its spherical harmonic coefficients $a_{\ell m}$ by
\begin{equation}
    M(\hat{\Omega}) = \sum\limits_{\ell m} a_{\ell m} Y_{\ell m}(\hat{\Omega})
    \qquad\text{and}\qquad
    a_{\ell m} = \int d^2\hat{\Omega}\, M(\hat{\Omega}) Y_{\ell m}^*(\hat{\Omega})\,,
\end{equation}
where $\hat{\Omega}=(\theta, \varphi)$ is the position on the celestial sphere.

A patch in the sky is selected by multiplying the map $M(\hat{\Omega})$ by the patch window $W(\hat{\Omega}, \hat{\Omega}_0)$ (patch centered at $\hat{\Omega}_0$). The latter can be either a simple mask (i.e.\ a step function, taking values 1 inside the patch and 0 outside) or a more general beam function with no sharp edges in pixel space. Hence, the position-dependent harmonic coefficients (i.e.\ the harmonic coefficients in the patch centered at $\hat{\Omega}_0$) are given by
\begin{equation}
    \label{eq:patch-alm}
    \tilde{a}_{\ell m}(\hat{\Omega}_0) 
     = \int d^2\hat{\Omega}\, M(\hat{\Omega}) W(\hat{\Omega},\hat{\Omega}_0) Y_{\ell m}^*(\hat{\Omega})
     = \sum\limits_{\ell_1 m_1} a_{\ell_1 m_1} K_{\ell \ell_1}^{m m_1}(\hat{\Omega}_0)\,,
\end{equation}
where we use the notation \textasciitilde{} to indicate that it is a patch quantity (with respect to the harmonic coefficient $a_{\ell m}$ for the full sky) and $K_{\ell \ell_1}^{m m_1}$ is the patch kernel defined by
\begin{equation}
    \label{eq:patch-kernel}
        K_{\ell_1 \ell_2}^{m_1 m_2} (\hat{\Omega}_0)
        \equiv \int d^2\hat{\Omega}\, W(\hat{\Omega},\hat{\Omega}_0)  Y_{\ell_2 m_2}(\hat{\Omega}) Y_{\ell_1 m_1}^*(\hat{\Omega})\,.
\end{equation}
The position-dependent power spectrum is then given by
\begin{equation}
    \label{eq:patch-cl}
    \begin{split}
        \tilde{C}_{\ell} (\hat{\Omega}_0)
        & = \frac{1}{f^W_\mathrm{sky}}\frac{1}{2\ell+1} \sum\limits_{\ell m}\tilde{a}_{\ell m}(\hat{\Omega}_0)\tilde{a}_{\ell m}^*(\hat{\Omega}_0) \\
        & =  \frac{1}{f^W_\mathrm{sky}}\frac{1}{2\ell+1} \sum\limits_{\ell_1 m_1} \sum\limits_{\ell_2 m_2}a_{\ell_1 m_1}a_{\ell_2 m_2}\sum\limits_{m} (-1)^m K_{\ell \ell_1}^{m m_1}(\hat{\Omega}_0) K_{\ell \ell_2}^{-m m_2}(\hat{\Omega}_0)\,,
    \end{split}
\end{equation}
where we used $(K_{\ell_1 \ell_2}^{m_1 m_2})^* = (-1)^{m_1 + m_2}K_{\ell_1 \ell_2}^{-m_1 -m_2}$.

\noindent The patch function can also be decomposed into spherical harmonics
\begin{equation}
    \label{eq:patch-decomposition}
    W(\hat{\Omega}, \hat{\Omega}_0) = \sum\limits_{\ell m} w_{\ell m}(\hat{\Omega}_0) Y_{\ell m}(\hat{\Omega})\,.
\end{equation} 
The local average value in a patch is then given by
\begin{equation}
    \label{eq:patch-avg}
    \begin{split}
        \tilde{\overbar{M}}(\hat{\Omega}_0) & = \frac{1}{4\pi f^W_\mathrm{sky}}\int d^2\hat{\Omega}\, M(\hat{\Omega}) W(\hat{\Omega},\hat{\Omega}_0)= \frac{1}{4\pi f^W_\mathrm{sky}}\sum\limits_{\ell m} \sum\limits_{\ell' m'} a_{\ell m} w_{\ell' m'}(\hat{\Omega}_0)\int d^2\hat{\Omega}\, Y_{\ell m}(\hat{\Omega}) Y_{\ell' m'}(\hat{\Omega})\\
        & = \frac{1}{4\pi f^W_\mathrm{sky}}\sum\limits_{\ell m} a_{\ell m} w^*_{\ell m}(\hat{\Omega}_0)\,,
    \end{split}
\end{equation}
where $f^W_\mathrm{sky}$ is the fraction of the sky covered by the patch and the last step is obtained using the orthogonality property of spherical harmonics. 

The integrated angular bispectrum is defined as the cross-correlation of the average $\tilde{\overbar{M}}$ and the position-dependent power spectrum $\tilde{C}_{\ell}$ of a patch, averaged over the whole sky: 
\begin{equation}
    \label{eq:ibisp-expectation}
        \ib_\ell \equiv \left\langle\frac{1}{4\pi}\int d^2\hat{\Omega}_0 \, \tilde{\overbar{M}}(\hat{\Omega}_0)\tilde{C}_{\ell} (\hat{\Omega}_0)\right\rangle\,.
\end{equation}
In practice, the integrated bispectrum of a map (simulation or observation) is computed by averaging over all the patches
\begin{equation}
    \label{eq:ibisp-obs}
    \ib_\ell^\mathrm{obs} = \frac{1}{N_\mathrm{patch}}\sum\limits_{\mathrm{patch}}
        \tilde{\overbar{M}}_\mathrm{patch}^\mathrm{obs}\tilde{C}_{\ell,\mathrm{patch}}^\mathrm{obs}\,,
\end{equation}
where $N_\mathrm{patch}$ is the total number of patches used to divide the sky and $\tilde{\overbar{M}}_\mathrm{patch}^\mathrm{obs}$ and $\tilde{C}_{\ell,\mathrm{patch}}^\mathrm{obs}$ are determined for each patch after multiplying the data map by the patch window. 

After substituting the position-dependent power spectrum (eq.\ \ref{eq:patch-cl}) and the mean patch value (eq.\ \ref{eq:patch-avg}) into eq.\ \eqref{eq:ibisp-expectation}, the expectation value of the integrated bispectrum becomes 
\begin{equation}
    \label{eq:ibisp-def}
    \begin{split}
        \ib_\ell& = \frac{1}{(4\pi)^2 (f^W_\mathrm{sky})^2} \frac{1}{2\ell+1} \sum\limits_{\ell_1 m_1} \sum\limits_{\ell_2 m_2} \sum\limits_{\ell_3 m_3} \langle a_{\ell_1 m_1}a_{\ell_2 m_2}a_{\ell_3 m_3}\rangle \\
        & \qquad \times \int d^2\hat{\Omega}_0 \, w^*_{\ell_3 m_3} (\hat{\Omega}_0) \sum\limits_{m} (-1)^m K_{\ell \ell_1}^{m m_1}(\hat{\Omega}_0) K_{\ell \ell_2}^{-m m_2}(\hat{\Omega}_0)\,.
    \end{split}
\end{equation}
The integrated bispectrum depends explicitly on the full angular bispectrum $\langle a_{\ell_1 m_1}a_{\ell_2 m_2}a_{\ell_3 m_3}\rangle$. The number of multipoles usually considered $\lmax\sim\mathcal{O}(10^3)$ makes the brute force computation of this summation totally out of reach, a common issue to all CMB bispectrum analyses. 

Assuming statistical isotropy, it is more convenient to work with a rotationally-invariant quantity called the angle-averaged bispectrum defined by\footnote{Note that the reduced bispectrum $B_{\ell_1 \ell_2 \ell_3} / h_{\ell_1 \ell_2 \ell_3}^2$, where $h_{\ell_1 \ell_2 \ell_3}$ is defined in eq.\ \eqref{eq:numtriangles} is also often used in the literature.} 
\begin{equation}
    \label{eq:angle-avg-bisp}
    B_{\ell_1 \ell_2 \ell_3} \equiv  \left\langle \int d^2\hat\Omega \,
     M_{\ell_1}(\hat\Omega) M_{\ell_2}(\hat\Omega) M_{\ell_3}(\hat\Omega)\right\rangle\,, 
    \quad\text{with}~~
    M_\ell(\hat\Omega)=\sum\limits_{m} a_{\ell m} Y_{\ell m}(\hat\Omega)\,.
\end{equation}
This angle-averaged bispectrum is related to the angular one by
\begin{equation}
    \label{eq:bisp-link}
    \langle a_{\ell_1 m_1}a_{\ell_2 m_2}a_{\ell_3 m_3}\rangle =  h_{\ell_1 \ell_2 \ell_3}^{-1} 
    \begin{pmatrix}
        \ell_1 &  \ell_2 &  \ell_3\\
        m_1 & m_2 & m_3
    \end{pmatrix}
    B_{\ell_1 \ell_2 \ell_3}\,,
\end{equation}
where the matrix is a Wigner-3$j$ symbol and $h_{\ell_1 \ell_2 \ell_3}$ is defined by
\begin{equation}
    \label{eq:numtriangles}
    h_{\ell_1 \ell_2 \ell_3}
    \equiv
    \sqrt{\frac{(2\ell_1+1)(2\ell_2+1)(2\ell_3+1)}{4\pi}}
    \begin{pmatrix}
      \ell _1 &\ell _2 &\ell _3\\
      0 & 0 & 0
    \end{pmatrix}\,.
\end{equation}
We also use the related Gaunt integral
\begin{equation}
    \label{eq:gaunt}
    \int d^2\hat{\Omega} \, Y_{\ell_1 m_1}(\hat{\Omega}) Y_{\ell_2 m_2}(\hat{\Omega}) Y_{\ell_3 m_3}(\hat{\Omega})
    = h_{\ell_1 \ell_2 \ell_3} \begin{pmatrix}
        \ell_1 &  \ell_2 &  \ell_3\\
        m_1 & m_2 & m_3
    \end{pmatrix}\,,
\end{equation}
and eq.\ \eqref{eq:patch-decomposition} to rewrite the patch kernel definition (eq.\ \ref{eq:patch-kernel}) as
\begin{equation}
        K_{\ell_1 \ell_2}^{m_1 m_2} (\hat{\Omega}_0)
 = (-1)^{m_1} \sum\limits_{\ell_3 m_3} w_{\ell_3 m_3}(\hat{\Omega}_0)\, h_{\ell_1 \ell_2 \ell_3} \begin{pmatrix}
            \ell_1 & \ell_2 & \ell_3\\
            -m_1 & m_2 & m_3
        \end{pmatrix}\,,
\end{equation}
After substituting this expression and eq.\ \eqref{eq:bisp-link} into the integrated bispectrum (eq.\ \ref{eq:ibisp-def}), we obtain
\begin{equation}
    \label{eq:ibisp-def2}
    \begin{split}
        \ib_\ell = 
        & \frac{1}{(4\pi)^2 (f^W_\mathrm{sky})^2} \frac{1}{2\ell+1} \sum\limits_{\ell_1 \ell_2 \ell_3} \frac{B_{\ell_1 \ell_2 \ell_3}}{h_{\ell_1 \ell_2 \ell_3}}  \sum\limits_{m_1 m_2 m_3} \begin{pmatrix}
            \ell_1 &  \ell_2 &  \ell_3\\
            m_1 & m_2 & m_3
        \end{pmatrix}\\
        &  \times \sum\limits_{m_4 m_5 m} (-1)^{m} 
        \begin{pmatrix}
            \ell &  \ell_1 &  \ell_4\\
            -m & m_1 & m_4
        \end{pmatrix}
        \begin{pmatrix}
            \ell &  \ell_2 &  \ell_5\\
            m & m_2 & m_5
        \end{pmatrix}
        \int d^2\hat{\Omega}_0 \, w^*_{\ell_3 m_3}(\hat{\Omega}_0) w_{\ell_4 m_4}(\hat{\Omega}_0) w_{\ell_5 m_5}(\hat{\Omega}_0)\,.
    \end{split}
\end{equation}
The remaining integral depends exclusively on the choice of patches and its dependence on several multipole numbers $m$'s still makes the integrated bispectrum impossible to compute in practice. Similarly to other bispectrum estimators, the solution comes under the form of separability of the integrated bispectrum in $\ell$-space, which we achieved using a class of simple azimuthally symmetric patches.

\subsection{Multipole space patches}
\label{sec:multipole-patch}

In this section (and in the rest of the paper), we study azimuthally symmetric patches having the following simple $m$-dependence:
\begin{equation}
    \label{eq:patch-harmonic}
    w_{\ell m}(\hat{\Omega}_0) = w_\ell \, Y_{\ell m}^*(\hat{\Omega}_0)\,.
\end{equation}
Using the addition theorem of spherical harmonics
\begin{equation}
    \label{eq:addition-theorem}
    \sum\limits_{m} Y_{\ell m}(\hat{\Omega}) Y_{\ell m}^*(\hat{\Omega}') 
    = \frac{2\ell + 1}{4\pi} P_\ell(\hat{\Omega}\cdot\hat{\Omega}')\,
\end{equation}
where $P_\ell$ is a Legendre polynomial, it is straightforward to obtain from eq.\ \eqref{eq:patch-decomposition} the simple and $m$-independent expression of the patch window function in real space
\begin{equation}
    \label{eq:patch-real}
    W(\hat{\Omega},\hat{\Omega}_0) = \sum\limits_{\ell} w_\ell \frac{2\ell + 1}{4\pi} P_\ell(\hat{\Omega}\cdot\hat{\Omega}_0)\,.
\end{equation}

With this type of patches, the remaining integral term in the integrated bispectrum expression (eq.\ \ref{eq:ibisp-def2}) becomes a Gaunt integral (eq.\ \ref{eq:gaunt}):
\begin{equation}
        \int d^2\hat{\Omega}_0 \, w^*_{\ell_3 m_3}(\hat{\Omega}_0) w_{\ell_4 m_4}(\hat{\Omega}_0) w_{\ell_5 m_5}(\hat{\Omega}_0) = (-1)^{m_3} w_{\ell_3} w_{\ell_4} w_{\ell_5} h_{\ell_3 \ell_4 \ell_5} 
        \begin{pmatrix}
            \ell_3 &  \ell_4 &  \ell_5\\
            -m_3  & m_4 & m_5
        \end{pmatrix}\,.
\end{equation}
Substituting this into eq.\ \eqref{eq:ibisp-def2} gives
\begin{equation}
    \label{eq:ibisp-def3}
    \begin{split}
        \ib_\ell 
        = &\frac{1}{(4\pi f^W_\mathrm{sky})^2} \frac{1}{2\ell+1} \sum\limits_{\ell_1 \ell_2 \ell_3 \ell_4 \ell_5} B_{\ell_1 \ell_2 \ell_3} \frac{h_{\ell \ell_1 \ell_4} h_{\ell \ell_2 \ell_5} h_{\ell_3 \ell_4 \ell_5}}{h_{\ell_1 \ell_2 \ell_3}} w_{\ell_3} w_{\ell_4} w_{\ell_5}  \\
        & \times \sum_{\substack{m_1 m_2 m_3\\m_4 m_5 m}} (-1)^{m+m_3} \begin{pmatrix}
            \ell_1 &  \ell_2 &  \ell_3\\
            m_1 & m_2 & m_3
        \end{pmatrix}
        \begin{pmatrix}
            \ell &  \ell_1 &  \ell_4\\
            -m & m_1 & m_4
        \end{pmatrix}
        \begin{pmatrix}
            \ell &  \ell_2 &  \ell_5\\
            m & m_2 & m_5
        \end{pmatrix}
        \begin{pmatrix}
            \ell_3 &  \ell_4 &  \ell_5\\
            -m_3  & m_4 & m_5
        \end{pmatrix}\,.
    \end{split}
\end{equation}
In this new expression, the $m$-dependent part (the second line) is only a summation of Wigner 3$j$-symbols, hence it is independent from the exact choice of patches. More importantly, this summation of Wigner $3j$-symbols reduces to a Wigner $6j$-symbol:
\begin{equation}
    \begin{split}
    \sum\limits_{\substack{m_1 m_2 m_3\\m_4 m_5 m}} (-1)^{m+m_3} \begin{pmatrix}
        \ell_1 &  \ell_2 &  \ell_3\\
        m_1 & m_2 & m_3
    \end{pmatrix}
    \begin{pmatrix}
        \ell &  \ell_1 &  \ell_4\\
        -m & m_1 & m_4
    \end{pmatrix}
    \begin{pmatrix}
        \ell &  \ell_2 &  \ell_5\\
        m & m_2 & m_5
    \end{pmatrix}
    \begin{pmatrix}
        \ell_3 &  \ell_4 &  \ell_5\\
        -m_3  & m_4 & m_5
    \end{pmatrix}\\
    = (-1)^{\ell_2 + \ell_4}
    \begin{Bmatrix}
        \ell_1 &  \ell_2 &  \ell_3\\
        \ell_5  & \ell_4 & \ell
    \end{Bmatrix}\,.           
\end{split}
\end{equation}
Substituting this and eq.\ \eqref{eq:numtriangles} into eq.\ \eqref{eq:ibisp-def3}, we obtain our final expression for the integrated bispectrum:
\begin{equation}
    \label{eq:ibisp}
        \ib_\ell 
        = \frac{1}{(4\pi)^3 (f^W_\mathrm{sky})^2} \sum\limits_{\ell_1 \ell_2 \ell_3 \ell_4 \ell_5} B_{\ell_1 \ell_2 \ell_3} \mathcal{F}_{\ell_1 \ell_2 \ell_3 \ell_4 \ell_5} w_{\ell_3} w_{\ell_4} w_{\ell_5}\,,
\end{equation}
where $\mathcal{F}_{\ell_1 \ell_2 \ell_3 \ell_4 \ell_5}$ is a shorthand notation for the part of the expression depending only on multipole numbers
\begin{equation}
    \label{eq:factor-F}
    \begin{split}
        \mathcal{F}_{\ell_1 \ell_2 \ell_3 \ell_4 \ell_5}
         =   &(-1)^{\ell_2 + \ell_4} (2\ell_4+1)(2\ell_5+1)  \\
        &\times \begin{pmatrix}
            \ell_1 &  \ell_2 &  \ell_3\\
            0  & 0 & 0
        \end{pmatrix}^{-1}
        \begin{pmatrix}
            \ell &  \ell_1 &  \ell_4\\
            0  & 0 & 0
        \end{pmatrix}
        \begin{pmatrix}
            \ell &  \ell_2 &  \ell_5\\
            0  & 0 & 0
        \end{pmatrix}
        \begin{pmatrix}
            \ell_3 &  \ell_4 &  \ell_5\\
            0  & 0 & 0
        \end{pmatrix}
        \begin{Bmatrix}
            \ell_1 &  \ell_2 &  \ell_3\\
            \ell_5  & \ell_4 & \ell
        \end{Bmatrix}\,.
    \end{split}
\end{equation}

As stressed out at the beginning of the section, the position-dependent power spectrum is by definition interesting as a tool to study the squeezed limit of the bispectrum, where one multipole number is small compared to the two others. With the following convenient choice of patch window function
\begin{equation}
    \label{eq:lw}
    w_\ell = 0 \text{ for }\ell > \ell_w \text{ with }  \ell_w\sim \mathcal{O}(10)\,,
\end{equation} 
it is possible to remove (almost all) non-squeezed terms from the sum in eq.\ \eqref{eq:ibisp} which implies a significant computational gain. With this choice, only the part of the bispectrum where $\ell_3 \leq \ell_w$ is considered in the summation. The multipoles $\ell_4$ and $\ell_5$ also have to be less than or equal to $\ell_w$. Moreover, recalling that the Wigner 3$j$-symbols are zero when the triangle inequality is not respected, the only non-vanishing terms of eq.\ \eqref{eq:ibisp} require that $\ell_1, \ell_2 \in [\ell - \ell_w, \ell + \ell_w]$. Therefore, the total number of terms in the sum for a given $\ell$ has been drastically reduced and is only of $\mathcal{O}(\ell_w^5)$ (compared to $\mathcal{O}(\lmax^5)$ before), which is fast to compute.\footnote{In addition of the superior bound $\ell_w$, one can also impose a minimal mutipole value below which the patch window function $w_\ell$ is zero. This will only decrease the definition intervals of $\ell_3, \ell_4$ and $\ell_5$.} 

For this broad class of patches, one can determine the exact expected integrated bispectrum from the theoretical bispectrum shape. Before using these results to analyze data from simulations and observations, we also need its expected covariance matrix, which can be computed following a similar method.

First, we recall the well-known expression of the bispectrum variance, valid in the usual weak non-Gaussianity case corresponding to CMB observations
\begin{equation}
    \label{eq:bispectrum-covariance}
    \begin{split}
    \langle B_{\ell_1 \ell_2 \ell_3} B_{\ell_1' \ell_2' \ell_3'} \rangle 
    = h_{\ell_1 \ell_2 \ell_3}^2 C_{\ell_1}C_{\ell_2}C_{\ell_3} &\left[ \delta_{\ell_1 \ell_1'}\delta_{\ell_2 \ell_2'}\delta_{\ell_3 \ell_3'} 
         + \delta_{\ell_1 \ell_2'}\delta_{\ell_2 \ell_1'}\delta_{\ell_3 \ell_3'} 
        + \delta_{\ell_1 \ell_1'}\delta_{\ell_2 \ell_3'}\delta_{\ell_3 \ell_2'} \right.\\
    & \left. \quad + \delta_{\ell_1 \ell_2'}\delta_{\ell_2 \ell_3'}\delta_{\ell_3 \ell_1'}
        + \delta_{\ell_1 \ell_3'}\delta_{\ell_2 \ell_2'}\delta_{\ell_3 \ell_1'} 
        + \delta_{\ell_1 \ell_3'}\delta_{\ell_2 \ell_1'}\delta_{\ell_3 \ell_2'}  \right].
    \end{split}
\end{equation}
It is then straightforward to obtain the covariance,
\begin{equation}
    \label{eq:ibisp_covariance}
    \begin{split}
        \ic_{\ell\ell'}  \equiv & \, \langle \ib_{\ell} \ib_{\ell'} \rangle\\
        = & \frac{1}{(4\pi)^6 (f^W_\mathrm{sky})^4}
        \sum\limits_{\substack{\ell_1 \ell_2 \ell_3 \ell_4 \ell_5 \\ \ell'_1 \ell'_2 \ell'_3 \ell'_4 \ell'_5}} \langle B_{\ell_1 \ell_2 \ell_3} B_{\ell_1' \ell_2' \ell_3'} \rangle \mathcal{F}_{\ell_1 \ell_2 \ell_3 \ell_4 \ell_5} \mathcal{F}_{\ell'_1 \ell'_2 \ell'_3 \ell'_4 \ell'_5} w_{\ell_3} w_{\ell_4} w_{\ell_5} w_{\ell'_3} w_{\ell'_4} w_{\ell'_5} \\
        = & \frac{1}{(4\pi)^6 (f^W_\mathrm{sky})^4} \sum\limits_{\substack{\ell_1 \ell_2 \ell_3 \ell_4 \ell_5 \\ \ell'_4 \ell'_5}}  C_{\ell_1}C_{\ell_2}C_{\ell_3} h_{\ell_1 \ell_2 \ell_3}^2 \mathcal{F}_{\ell_1 \ell_2 \ell_3 \ell_4 \ell_5} w_{\ell_3} w_{\ell_4} w_{\ell_5} w_{\ell'_4} w_{\ell'_5} \\
        & \mkern-68mu \times \left[w_{\ell_3} (\mathcal{F}_{\ell_1 \ell_2 \ell_3 \ell'_4 \ell'_5} + \mathcal{F}_{\ell_2 \ell_1 \ell_3 \ell'_4 \ell'_5}) + w_{\ell_2} (\mathcal{F}_{\ell_1 \ell_3 \ell_2 \ell'_4 \ell'_5} + \mathcal{F}_{\ell_3 \ell_1 \ell_2 \ell'_4 \ell'_5}) + w_{\ell_1} (\mathcal{F}_{\ell_3 \ell_2 \ell_1 \ell'_4 \ell'_5} + \mathcal{F}_{\ell_2 \ell_3 \ell_1 \ell'_4 \ell'_5})\right],
    \end{split}
\end{equation}
where the last step is obtained after summing over $\ell'_1$, $\ell'_2$ and $\ell'_3$. Note that the terms in $w_{\ell_1}$ and $w_{\ell_2}$ on the last line are zero if $\ell > 2\ell_w$ (thus for most of the $\ell$'s considered). Note also that the sum is separable in $(\ell_4, \ell_5)$ and $(\ell'_4,\ell'_5)$, hence in practice there are only $\mathcal{O}(\ell_w^5)$ different terms to evaluate, as for the integrated bispectrum (eq.\ \ref{eq:ibisp}).

\section{The integrated bispectrum estimator}
\label{sec:ib-estimator}

We now describe a data-analysis procedure based on the analytical results of the previous section. First, we build a new estimator of the bispectrum amplitude, $\fnl$, using only the integrated bispectrum.
Then we discuss the necessary corrections to study observational data, with a focus on the standard but non-trivial linear correction to the bispectrum. Finally, we detail the numerical implementation of this integrated bispectrum estimator. Note that the methodology we follow in this section is inspired by the standard bispectrum estimators, thus we refer the reader to the literature available on this topic (see \cite{Liguori:2010hx} for a review and references therein for more details).

\subsection{$\hat{f}_\mathrm{NL}$ estimator}
\label{sec:fNL}

Compared to the bispectrum and its $\mathcal{O}(\ell_\mathrm{max}^3 / 6)$ different multipole triplets ($\ell_1, \ell_2, \ell_3$), the integrated bispectrum, which only has $\lmax$ configurations, is already a useful compression of the bispectral data. Moreover, this remaining $\ell$-dependence is useful to distinguish between different squeezed shapes in a very visual way (see figure \ref{fig:Tyy-signaltonoise} for example).

However, to deal with the very low level of non-Gaussianity in (cleaned) CMB observations, the usual method is to compress the data even further into a single parameter $\fnl$ which represents the amplitude at which a theoretical bispectrum is present in the data. Then, for each given shape the corresponding parameter $\fnl$ is estimated from the data by performing a weighted sum over all the multipole triplets of the ratio of the observed bispectrum to a given theoretical shape. A simple, standard derivation of these weights \cite{Komatsu:2001rj} starts from the assumption that the bispectrum configurations obey a Gaussian distribution to write the associated $\chi^2$. One then minimizes it to obtain the maximum likelihood estimator. 
Here we apply the same procedure to the integrated bispectrum, starting with the $\chi^2$:
\begin{equation}
    \label{eq:chi-fnl}
    \chi^2 = \sum\limits_{\ell\ell'} (\ib_\ell^\mathrm{obs} - \fnl \ib_\ell^\mathrm{th})(\ic^{-1})_{\ell\ell'} (\ib_{\ell'}^\mathrm{obs} - \fnl \ib_{\ell'}^\mathrm{th})\,.
\end{equation}
Minimizing the $\chi^2$ leads to the usual form of the estimator:%
\footnote{The estimator $\hat{f}_\mathrm{NL} = \frac{\sum\limits_{\ell} \ib_\ell^\mathrm{obs} \ib_\ell^\mathrm{th}/\ic_{\ell\ell}}{\sum\limits_{\ell} (\ib_\ell^\mathrm{th})^2 /\ic_{\ell\ell}}$ with an expected error bar 
$\sigma(\hat{f}_\mathrm{NL})=\frac{\sqrt{\sum\limits_{\ell\ell'} \ib_\ell^\mathrm{th}\ic_{\ell\ell'} \ib_{\ell'}^\mathrm{th}/\ic_{\ell\ell}/\ic_{\ell'\ell'}}}{\sum\limits_{\ell} (\ib_\ell^\mathrm{th})^2 /\ic_{\ell\ell}}$\,, is a simpler alternative considering only the diagonal part of the covariance. Its main advantage is that it does not require to invert the covariance matrix $\ic_{\ell\ell}$.
For all the tests and analyses of the following sections, it yields results which are similar to those obtained with eq.\ \eqref{eq:fnl}. However, we do not guarantee this is a general result.}
\begin{equation}
    \label{eq:fnl}
    \hat{f}_\mathrm{NL} = \frac{\langle \ib^\mathrm{obs}, \ib^\mathrm{th} \rangle}{\langle \ib^\mathrm{th}, \ib^\mathrm{th} \rangle},\qquad
    \text{where}\quad
    \langle \ib^A, \ib^B \rangle = \sum\limits_{\ell\ell'} \ib_\ell^A (\ic^{-1})_{\ell\ell'} \ib_{\ell'}^B\,,
\end{equation}
with an expected error bar $\sigma(\hat{f}_\mathrm{NL})=1/\sqrt{\langle \ib^\mathrm{th}, \ib^\mathrm{th}\rangle}$. Assuming that we use the multipole space patches described in section \ref{sec:multipole-patch}, the covariance $\ic_{\ell\ell'}$ is non-diagonal because from eq.\ \eqref{eq:ibisp_covariance} it is also non-zero if $\ell$ and $\ell'$ are close enough ($|\ell-\ell'|\leq 2\ell_w$).

The estimator above is for single shape analysis. For joint analysis of more shapes  (denoted by the indices $A,B$) we have 
\begin{equation}
    \label{eq:fnl-joint}
    \hat{f}_\mathrm{NL}^A = \sum\limits_B (F^{-1})_{AB}{\langle \ib^\mathrm{obs}, \ib^B\rangle}, \qquad
    F_{AB} = \langle \ib^A, \ib^B \rangle\,,
\end{equation}
where $F_{AB}$ is the Fisher matrix. This estimator has an expected error bar $\sigma(\hat{f}_\mathrm{NL}^A)=\sqrt{(F^{-1})_{AA}}$. One can also use the Fisher matrix elements to evaluate a correlation coefficient between two different shapes. Again this is a standard definition in bispectrum analysis taking the familiar form
\begin{equation}
    \label{eq:correlation-coefficient}
    c_{AB} = \frac{F_{AB}}{\sqrt{F_{AA}F_{BB}}}\,.
\end{equation}
Note that these correlation coefficients can differ from the usual bispectrum ones because only the squeezed part of the bispectrum is considered.

Some shapes, like those arising from secondary anisotropies in the CMB (e.g. ISW-lensing bispectrum), have an amplitude fixed by cosmological parameters. 
Often these shapes play in the analysis the role of spurious contaminant of the primordial NG signal, and their effect must be marginalized out.
However, if we have an accurate measurement of cosmological parameters, we can also proceed in a simplified way by assuming a best-fit cosmology and computing the expected bias, produced by the spurious shape on the shape of interest. For a shape $A$ contaminated by another known shape $B$, $\fnl^A$ becomes
\begin{equation}
    \label{eq:fnl-bias}
    \hat{f}_\mathrm{NL}^A = \frac{1}{F_{AA}}\langle \ib^\mathrm{obs}, \ib^A \rangle 
    - \frac{F_{AB}}{F_{AA}} \fnl^B \,,
\end{equation}
where the first part is just the independent estimation of $\fnl^A$ and the second part is the expected bias on $\hat{f}_\mathrm{NL}^A$ due to shape $B$.

We recall that all the expressions given in this section have the same form when working with an optimal bispectrum estimator, after applying the following substitution:
\begin{equation}
   \langle B^A, B^B \rangle=\sum\limits_{\ell_1\leq\ell_2\leq\ell_3}\frac{B_{\ell_1\ell_2\ell_3}^AB_{\ell_1\ell_2\ell_3}^B}{V_{\ell_1\ell_2\ell_3}}\,,
\end{equation}
where $V_{\ell_1\ell_2\ell_3}$ is the variance of the bispectrum (eq. \ref{eq:bispectrum-covariance}).

\subsection{Corrections for observational data}
\label{sec:ib_obs}

When considering realistic datasets, several effects have to be taken into account, which complicate the theoretical picture given in the previous section. In order to compare observations to theoretical expectations, we need to extend our estimator by including instrumental effects and corrections for partial sky coverage and non-stationary noise.

\subsubsection{Instrument}
\label{sec:instrument}

If we consider CMB, or more in general intensity mapping data, we can include the characteristics of the instrument (i.e.\ angular resolution, noise) in the power spectrum and the bispectrum, using the following modifications
\begin{equation}
    \label{eq:beam-noise}
    C_\ell \rightarrow b_\ell^2 C_\ell + n_\ell \qquad\text{and}\qquad
    B_{\ell_1 \ell_2 \ell_3} \rightarrow b_{\ell_1}b_{\ell_2}b_{\ell_3}B_{\ell_1 \ell_2 \ell_3} \, ,
\end{equation}
where $b_\ell$ is the beam transfer function and $n_\ell$ is the noise power spectrum.%
\footnote{$b_\ell$ is often approximated by a Gaussian beam $b_\ell=\exp\left[-\frac{1}{2}\frac{\ell(\ell+1)\theta_\mathrm{FWHM}^2}{8\ln2}\right]$, fully characterized by $\theta_\mathrm{FWHM}$.}
Moreover, $b_\ell$ has often to be multiplied by the pixel window function to take into account pixelization effects. At the integrated bispectrum level, it is then sufficient to substitute these corrections into eqs.\ \eqref{eq:ibisp} and \eqref{eq:ibisp_covariance}. 

In practice, some parts of the sky are observed more times than others (depending on the scanning pattern of the satellite), leading to an anisotropic distribution of the noise on the celestial sphere. Hence, each patch has its own distinct level of noise affecting its local power spectrum (larger for a noisier patch) while the local average is untouched (it is of cosmological origin only). The resulting spurious correlations can be misinterpreted as a large integrated bispectrum (or more generally as a large squeezed bispectrum), increasing the variance of the estimator. The solution to this problem is the so-called linear correction, which we will detail in section \ref{sec:linear-correction}.

\subsubsection{Mask}
\label{sec:mask}

Both CMB and LSS datasets are characterized by incomplete sky coverage, either because only part of the sky is actually observed, or because foreground contamination imposes to mask a non-negligible part of it.

The missing power from the masked/non-observed regions induces a multiplicative bias on the observed quantities:
\begin{equation}
    C_\ell^{\mathrm{masked}} = f_\mathrm{sky} C_\ell^{\mathrm{unmasked}}, \quad
    B_{\ell_1 \ell_2 \ell_3}^{\mathrm{masked}} = f_\mathrm{sky} B_{\ell_1 \ell_2 \ell_3}^{\mathrm{unmasked}} \quad\text{and}\quad
    \ib_\ell^{\mathrm{masked}} = f_\mathrm{sky} \ib_\ell^{\mathrm{unmasked}},
\end{equation} 
the so-called $f_\mathrm{sky}$ approximation, where $f_\mathrm{sky}$ is the fraction of the sky that is left unmasked (see \cite{Komatsu:2001wu} for the derivation of the first two equations while the last one is obtained by substituting the corrected bispectrum into eq.\ \ref{eq:ibisp}).

Moreover, the abrupt step at the edge of the mask in real space produces power leaking between multipoles.
The solution is to smooth these edges, which can be done by employing several different existing methods (see \cite{Gruetjen:2015sta} for more details). In this paper, we use a diffuse inpainting approach, similar to the one adopted in the \planck\ non-Gaussianity analyses \cite{Akrami:2019izv}. The masked region is filled in by the average of the map, and then, iteratively, every masked pixel is given the average value of its neighbours (2000 iterations for \planck\ maps are sufficient).

A final, important effect of partial sky coverage is that it breaks rotational invariance. As in the case of the anisotropic spatial distribution of the noise, the solution usually comes under the form of a linear correction to the bispectrum.

\subsubsection{Linear correction}
\label{sec:linear-correction}

We described the qualitative effects of breaking rotational invariance with the example of the anisotropic noise in section \ref{sec:instrument}. At the quantitative level, the bispectrum variance expression (eq.\ \ref{eq:bispectrum-covariance}) is only valid for an isotropic observation of the sky and a more general formulation contains additional terms increasing it significantly (see \cite{2012ApJ...755...19D} for a detailed computation). However, it is possible to use a slightly modified version of the observed bispectrum composed of the usual cubic part ($a_{\ell_1 m_1} a_{\ell_2 m_2} a_{\ell_3 m_3}$) and of a linear correction term ($a_{\ell m}$). This cancels out all the extra terms in the variance to recover an expression of the same form as eq.\ \eqref{eq:bispectrum-covariance} without adding any biases to the measurements (the linear term is zero on average). This solution was first implemented and used in \cite{Creminelli:2005hu}.

Let us now see how we can specifically build the linear correction for the integrated bispectrum estimator. To fix ideas, let us still consider the case of an anisotropic noise pattern in a CMB experiment, even though the final result will be completely general. The goal is to remove the excess of variance due to the correlations between the large-scale and the small-scale fluctuations of the noise, induced by the scanning pattern of the satellite. This effect can be computed using a Monte-Carlo average over many simulations having the same characteristics as the observational data. The idea is to estimate the correlations between the mean values of the patches of the actual data set (i.e.\ observed large-scale modes, including both signal and noise) and the {\em noise} position-dependent power spectrum, determined from realistic simulations (i.e.\ small-scale noise fluctuations). If there is a mask, the method is the same. However, the simulations must include in this case also the cosmological signal, and not just the noise. The linear, mean-field correction for the integrated bispectrum then takes the form
\begin{equation}
    \label{eq:linear-correction}
    \ib_\ell^\mathrm{lin} = \frac{1}{N_\mathrm{patch}}\sum\limits_{\mathrm{patch}}
        \tilde{\overbar{M}}_\mathrm{patch}^\mathrm{obs}\tilde{C}_{\ell,\mathrm{patch}}^\mathrm{MC}\,,
\end{equation}
where $\tilde{C}_{\ell,\mathrm{patch}}^\mathrm{MC}$ position-dependent power obtained by averaging over a set of simulations (typically $\mathcal{O}(100)$), with the same characteristics as the data set. The observed integrated bispectrum given in eq.\ \eqref{eq:ibisp-obs} should then be substituted by
\begin{equation}
    \label{eq:ibisp-obs+lin}
    \ib_\ell^\mathrm{obs}\rightarrow\ib_\ell^\mathrm{obs}-\ib_\ell^\mathrm{lin}\,.
\end{equation}

\subsection{Implementation of the integrated bispectrum}
\label{sec:implementation}

The integrated bispectrum is clearly much easier and more straightforward to measure than the full bispectrum (which is the reason for its introduction in first place). 
If, besides measuring the integrated bispectrum of data and simulations, we are also interested in determining amplitude parameters via template fitting, an additional advantage is that the observational (determining the integrated bispectrum of the studied map) and the theoretical (computing integrated bispectrum templates) parts of the fitting procedure are independent and can be performed separately. This makes it easy to check for additional shapes for example (a similar advantage is shared by binned and modal methods, for full bispectrum estimation).

\begin{figure}
    \centering
    \includegraphics[width=0.99\linewidth]{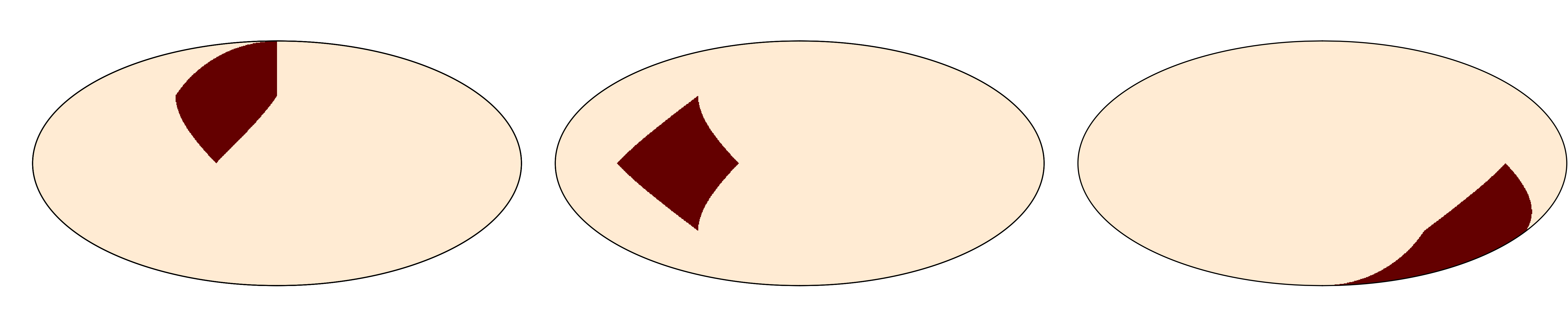}
    \includegraphics[width=0.99\linewidth]{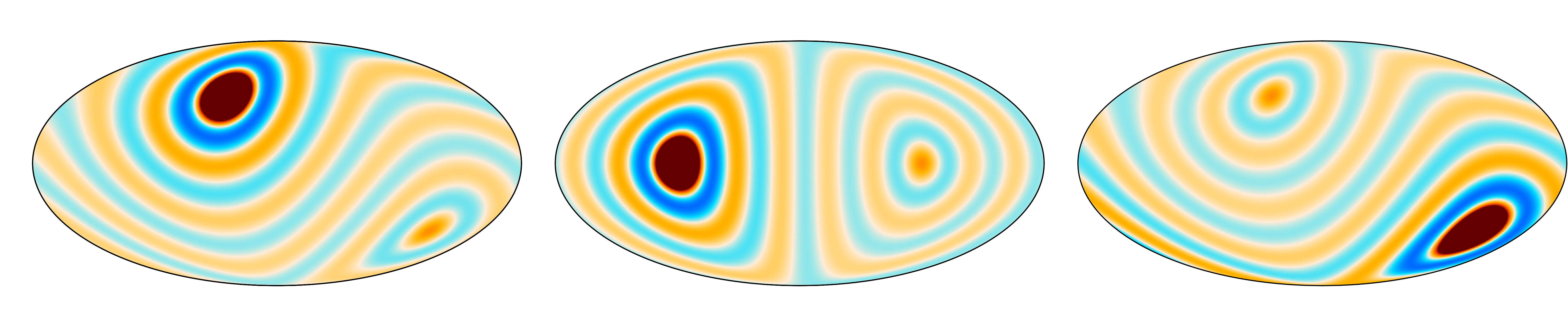}
    \includegraphics[width=0.99\linewidth]{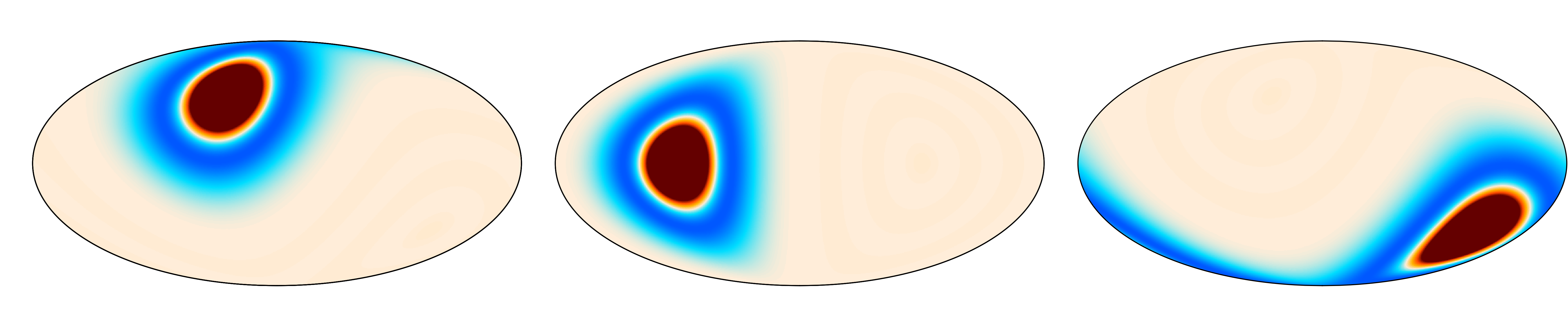}
    \includegraphics[width=0.6\linewidth]{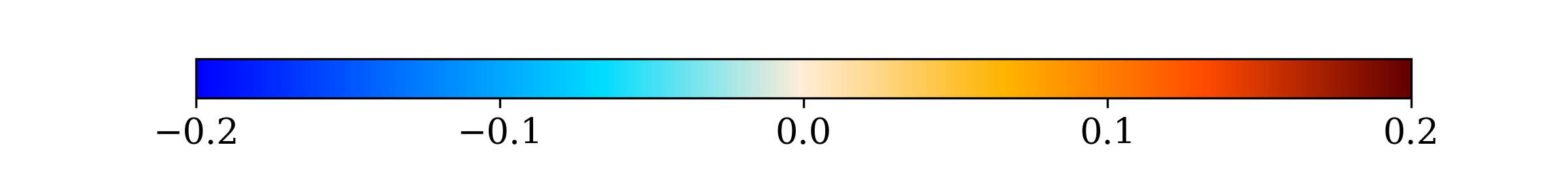}
    \caption{The different sets of patches used in this paper. The top panels shows three HEALPix patches for a sky divided in 12 areas (they are equal to 1 in the red areas and 0 otherwise). The middle and bottom panels represents multipole space patches as defined in section \ref{sec:multipole-patch}, centered at the same positions as the HEALpix patches shown above and being non-zero everywhere. The middle plots are step function patches ($w_\ell = 1$ for $\ell \leq \ell_w = 10$) which are used extensively in section \ref{sec:data}. The bottom plots are Mexican needlet patches defined in appendix \ref{eq:mexican-needlets}. Note that the absolute scale here has no physical meaning.}
    \label{fig:patches}  
\end{figure}

Concerning the observational component of the code, the first step is to generate the patches. 
One possibility would be to use HEALPix pixelization \cite{Gorski:2004by}, which, thanks to its hierarchical nature, offers a simple way to divide the celestial sphere in a set of equal-sized areas by simply taking the (large) pixels of a very low resolution maps.%
\footnote{\url{https://healpix.sourceforge.io}}
For example to generate 192 patches, make a map with the resolution $\nside=4$ and a zero value for each of its 192 pixels. Then, fix the value of a pixel to one and upgrade the resolution to the same as the studied data map (e.g., $\nside = 2048$ for \planck) to obtain one patch. Repeat the same process for each pixel of the low resolution map to obtain the full set. However, these patches are not suited for analytical computations of the integrated bispectrum because they do not have a simple expression in multipole space (they do not allow removing $m$-dependence in the final integrated bispectrum expression). This is why, except for some results on CMB simulations (both Gaussian and non-Gaussian) presented in appendix \ref{sec:healpix-patch}, we will not use this approach in the rest of the paper and will instead focus on the azimuthally-symmetric, multipole space patches introduced in section \ref{sec:multipole-patch}.
Their simple definition in multipole space (eq.\ \ref{eq:patch-harmonic}) can be translated into real space maps using eq.\ \eqref{eq:patch-real}, requiring to compute Legendre polynomials for every pixel of the map. This takes significantly longer than the HEALpix patches approach, but the computational time remains still very reasonable. An important point to notice is that multipole space patches have non-zero values everywhere in the sky (the patch window is not a step function in pixel space), thus there is not a unique number of different patches leading to a complete, isotropic coverage of the sky.
A good method to obtain such a coverage is to use a number of multipole patches again corresponding to the number of pixels in a low resolution HEALPix map (we will use $\nside=4$ or $8$ in our analysis, corresponding to $192$ or $768$ pixels) and to center each of them at the center of a corresponding HEALPix pixel patch. Computing 192 patches in this way is typically as long as the following step of the pipeline, which is that of computing the power spectrum in each patch. Therefore, it is important to save real space maps for each patch, when simultaneously analyzing many data maps. In figure \ref{fig:patches}, we plot a few examples of the two types of patches we have just discussed. Note that the HEALPix patches, which are shown in the top panels, are larger than the ones we used for the analysis presented in appendix \ref{sec:healpix-patch} (the sky is divided into 12 equal-sized areas instead of 192 for visibility). For the azimuthally-symmetric patches, we consider two possible different choices of the patch window. The plots in the middle correspond to $\ell$-space step function patches  ($w_\ell = 1$ for $\ell \leq 10$ and $0$ otherwise). The bottom ones are Mexican needlets, which we discuss briefly in appendix \ref{eq:mexican-needlets}. Both types of multipole space patches have non-zero values everywhere in the sky. They are largest around their center and present damped oscillations around 0, further away from it. 
One of the general advantages of Mexican needlet patches is their localization in real space (among others). However, our goal is to look at the squeezed configurations of the bispectrum only. Therefore, strong localization in real space is not necessary and in the end we decide to adopt the step-function windows in all the analyses presented in section \ref{sec:data}.

After applying the observational mask (with the inpainting procedure, see section \ref{sec:ib_obs}), the main step of the analysis is as simple as computing the monopole and the power spectrum of our masked map, multiplied by a patch function, and repeating the procedure for all previously determined patches. Hence, the amount of required computations is that of $N_\mathrm{patches}$ (number of patches) evaluations of a power spectrum, using the HEALPix \texttt{anafast} tool. The integrated bispectrum is then finally obtained by averaging the products between the monopole and the power spectrum in each patch.

Whenever a mask is used or the noise pattern is not isotropic, the linear correction described in section \ref{sec:linear-correction} (eq.\ \ref{eq:linear-correction}) is necessary. To compute it, we use Gaussian simulations (we find that typically 100 of them is sufficient). The power spectrum in each patch of each simulation is then evaluated in the same way as described in the previous paragraph (this requires the evaluation of $N_\mathrm{simulations}\times N_\mathrm{patches}$ power spectra). Then the correction term can be computed and subtracted from the observed integrated bispectrum. 

After estimating the integrated bispectrum of the data, we might be in some cases interested in fitting specific theoretical shapes to observationally determine the corresponding amplitude parameters. In this case, theoretical templates of the integrated bispectrum for any given shape can be built via a direct implementation of eqs.\ \eqref{eq:ibisp} and \eqref{eq:ibisp_covariance} (which assume azimuthally symmetric patches). This requires precomputed power spectra $C_\ell$ (in our analysis we use CAMB\footnote{\url{http://camb.info/}} for this \cite{Lewis:1999bs}) and bispectra $B_{\ell_1\ell_2\ell_3}$.
The most intensive part of the computation is related to the calculation of the different Wigner symbols, which we determine using the standard recursion algorithms of \cite{doi:10.1063/1.522426}. For $\lmax=3000$ and $\ell_w=10$ (like in figure \ref{fig:patches}), it takes only a few seconds to compute a theoretical integrated bispectrum with its covariance on a simple laptop.

\section{Applications}
\label{sec:data}

In this section, we apply the pipeline described in section \ref{sec:implementation} to CMB data analysis. The full CMB bispectrum has been thoroughly studied, both theoretically and observationally, making it an ideal observable for a first application of the 2D integrated angular bispectrum estimator. Comparisons with results from full bispectrum estimators will allow us to evaluate the performance of the integrated bispectrum and to show in which areas it can be most useful. We stress however again that all the results obtained and discussed so far (both the analytic formulae and their pipeline implementation) are fully general. Therefore they can be applied to any 2D cosmological field on the sphere, and not just to CMB fluctuations.

In the applications we are going to present, we consider different CMB bispectrum shapes, peaking in the squeezed limit. First, we look at the standard local and ISW-lensing shapes. We use formulae obtained in the previous sections to numerically compute the angular integrated bispectrum templates arising from these shapes and compare them to estimates from NG CMB simulations. These constitute our validation tests of the new method. 
We then discuss in detail the integrated bispectrum related to another squeezed shape, namely the bispectrum due to correlations between the ISW and tSZ effects. These three shapes are finally included in a new analysis of the \planck\ CMB temperature data, which is independent from the standard methods.
We then briefly discuss the integrated bispectrum of thermal dust emission and we finish by a section on forecast predictions.

\subsection{Local shape}
\label{sec:local}

Among the shapes peaking in squeezed configurations, the most studied is of primordial origin and is called the local bispectrum. This shape is produced for example in multiple field inflation, because of the interactions between isocurvature and adiabatic perturbations on superhorizon scales. Originally, it was introduced by expanding at second order the primordial gravitational potential as
\begin{equation}
    \label{eq:local-potential}
    \Phi(\mathbf{x}) = \Phi_L(\mathbf{x}) 
    + \fnlloc (\Phi_L^2(\mathbf{x}) - \langle\Phi_L^2\rangle)\,,
\end{equation}
where the first term $\Phi_L$ denotes the linear (and Gaussian) part of the perturbation and the second term is the non-linear correction part whose amplitude is the parameter $\fnlloc$.
The locality of this formula in real space translates to correlations between very different modes in harmonic space and it actually peaks on squeezed triangles ($\ell_1 \ll \ell_2\approx\ell_3$). Therefore, the integrated bispectrum is well-adapted for its study, as we will verify in this section.

The local angle-averaged bispectrum can be written as an integral of a (look-back) conformal distance \cite{Komatsu:2001rj}:
\begin{equation}
    B_{\ell_1 \ell_2 \ell_3}^\mathrm{local} = 2 h_{\ell_1 \ell_2 \ell_3}^2 
    \int_0^\infty r^2\diff r\, \left[\alpha_{\ell_1}\beta_{\ell_2}\beta_{\ell_3} + \alpha_{\ell_2}\beta_{\ell_1}\beta_{\ell_3} + \alpha_{\ell_3}\beta_{\ell_1}\beta_{\ell_2}\right]\,,
\end{equation}
where
\begin{equation}
    \alpha_\ell(r) \equiv \frac{2}{\pi} \int_0^\infty k^2\diff k\, \Delta_\ell(k) j_\ell(kr),\qquad
    \beta_\ell(r) \equiv \frac{2}{\pi} \int_0^\infty k^2\diff k\, P_\Phi(k) \Delta_\ell(kr) j_\ell(k)\, \,.
\end{equation}
In these expressions, $P_\Phi(k)$ is the primordial power spectrum of the gravitational potential, the $j_\ell$ are spherical Bessel functions and the $\Delta_\ell$ are the radiation transfer functions which describe the linear evolution of the perturbations.

We now perform a series of tests on CMB simulations containing a known level of local non-Gaussianity, in order to validate the integrated bispectrum method and its efficiency at constraining $\fnlloc$. The NG maps are produced using the method described in \cite{Liguori:2003mb, 2009ApJS..184..264E}.

The goal of our first test is to check the agreement between the theoretical results of section \ref{sec:multipole-patch} and the same quantities determined directly from two sets of 1000 CMB maps, one Gaussian ($\fnlloc=0$) and the other non-Gaussian ($\fnlloc=50$), both identical at the linear level. We use low resolution maps ($\nside=256$) with $\lmax=500$. These maps contain no noise, no beam (except a pixel window function) and are analyzed in the full sky case (no mask). The integrated bispectrum is determined using 192 patches defined by a step function in multipole space ($w_\ell=1$ for $\ell \leq 10$ and 0 otherwise).
In appendix \ref{sec:patch-number}, we justify this choice of number of patches by showing that 192 is optimal here. The maps of the patches (in real space) are then generated by applying eq.\ \eqref{eq:patch-real} (see the middle line of figure \ref{fig:patches}).

\begin{figure}
    \centering
    \includegraphics[width=\textwidth]{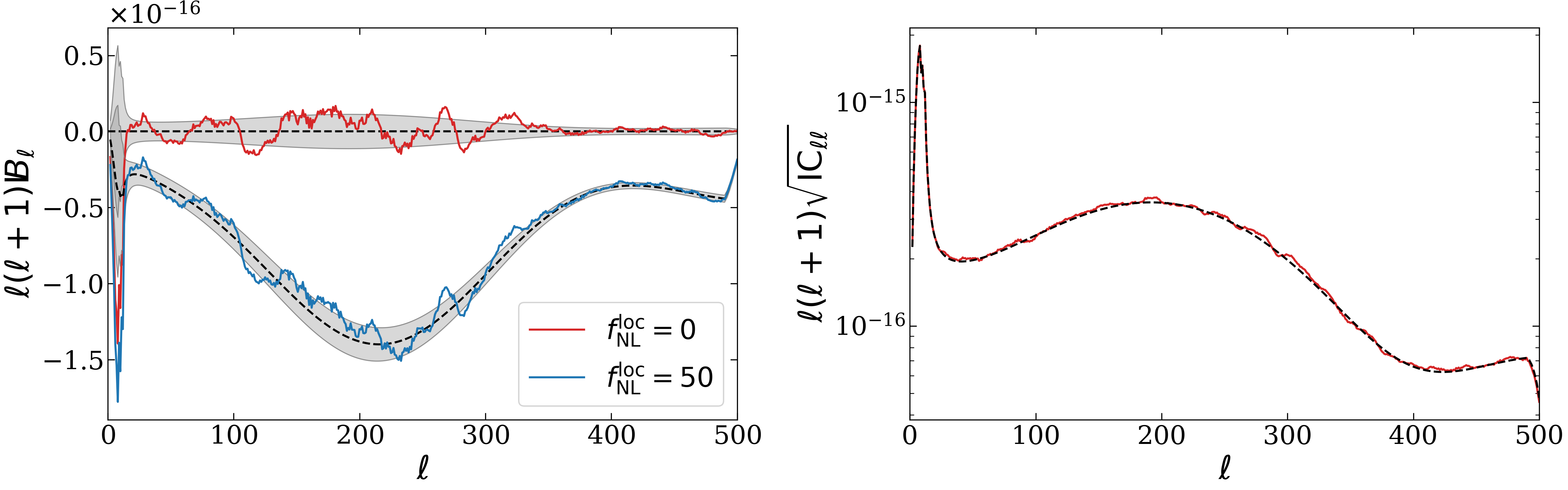}
    \caption{The integrated bispectra from CMB simulations ($\nside=256$). On the left, the averaged integrated bispectra from 1000 Gaussian maps ($\fnlloc=0$) in red and from 1000 non-Gaussian maps ($\fnlloc=50$) in blue. The black dashed lines correspond to their respective theoretical predictions computed using eq.\ \eqref{eq:ibisp} with their expected standard error (68\% CL) in grey, computed with eq.\ \eqref{eq:ibisp_covariance}. On the right, the standard deviation from the 1000 Gaussian maps in red and its theoretical counterpart in black (dashed line). By definition, the integrated bispectrum is similar to a power spectrum, thus it is convenient to multiply it by the standard factor $\ell(\ell+1)$ for a better visual representation. Moreover, the local integrated bispectrum actually has the same scaling as the primordial power spectrum (except at very low $\ell$'s). Note that the scale (vertical axis) is logarithmic in the right panel.}
    \label{fig:loc-256}  
\end{figure}

In figure \ref{fig:loc-256} we show the averaged integrated bispectra of the two sets of maps and their theoretical counterparts.
In figure \ref{fig:loc-256}, we also show, for each multipole, the standard deviation of the integrated bispectrum derived from the Gaussian maps (in the non-Gaussian case, the variance increases). In both plots, the results determined from the maps are in perfect agreement with the theoretical expectations.
Moreover, applying the estimator (eq.\ \ref{eq:fnl}) yields the mean values of $\fnlloc = -0.75 \pm 0.68$ and $\fnlloc = 48.78 \pm 0.78$ for the Gaussian and non-Gaussian maps respectively. In these results, we show the standard error on the mean of $\fnlloc$, estimated from 1000 maps. The corresponding $\fnlloc$ standard deviation is 21.3 in the Gaussian case (the theoretical Fisher error bar is also 21.3). These are 25\% larger than the optimal results, achievable by a full bispectrum estimator at $\lmax=500$ ($\sigma_\text{opt}(\fnlloc)=16.5)$.
For the non-Gaussian maps, the larger error bar is due to the fact that the weak non-Gaussianity approximation starts to break down because $\fnlloc=50$. The consequence is that the inverse covariance weights of the estimator (eq.\ \ref{eq:fnl}), which are derived using this approximation (otherwise eq.\ \eqref{eq:bispectrum-covariance} is not valid), become suboptimal (for a discussion on this effect see \cite{Creminelli:2006gc,Liguori:2007sj}. 

These different results prove the validity of the method in this ideal, low resolution case, motivating us to study next a more realistic situation.

The second test is closer to a \planck-like situation, to check the different corrections introduced in section \ref{sec:ib_obs}. Here we use high-resolution maps ($\ell_\mathrm{max} \sim 2000$), partially masked and with non-negligible instrumental effects (beam, noise). 

\begin{figure}
    \centering
    \includegraphics[width=\textwidth]{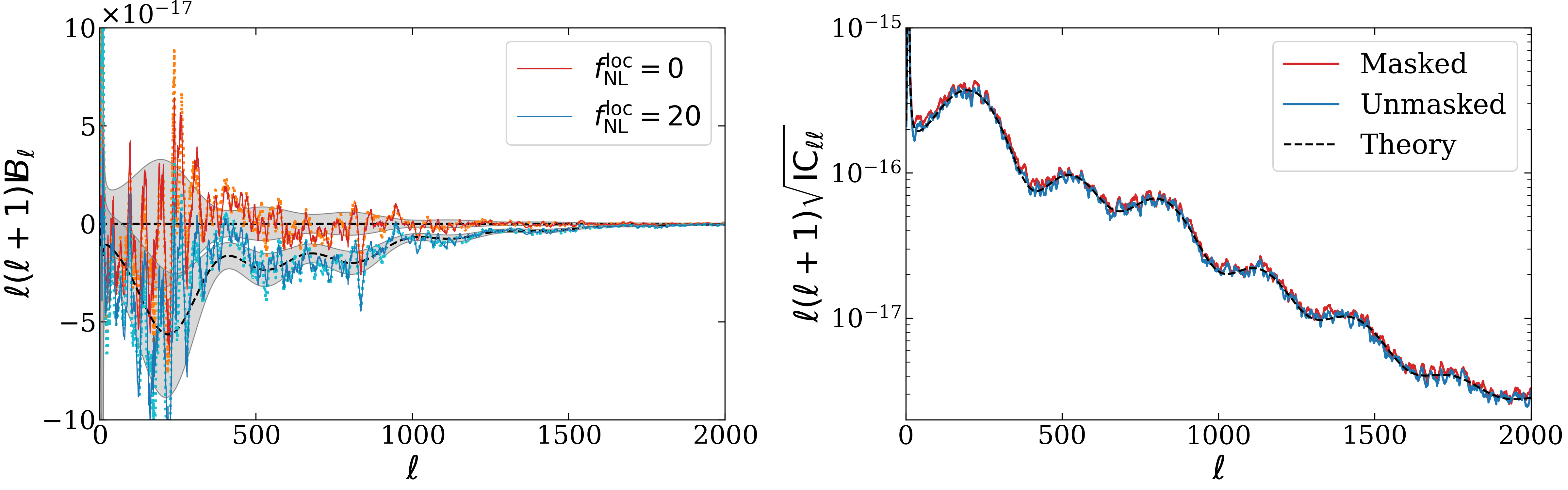}
    \caption{Similar to figure \ref{fig:loc-256}, with sets ($\fnlloc=0$ and $\fnlloc=20$) of 100 high resolution maps ($\nside = 2048$). The analyses are performed in the full sky and masked (\planck\ 2018 common mask, $\fsky=0.78$) cases, in orange and red for the Gaussian maps respectively and cyan and blue for the non-Gaussian ones on the left panel. On the right, the standard deviation of the Gaussian maps in green (full-sky) and red (masked).}
    \label{fig:loc-2048}  
\end{figure}

At high resolution the computation time increases considerably, hence we only use 100 maps with $\fnlloc=0$ and $\fnlloc=20$ (to avoid breaking the weak non-Gaussianity approximation too strongly). These maps include the effects of a Gaussian beam with a 5 arcmin FWHM and a noise power spectrum similar to the \planck\ one (for now, with an isotropic spatial distribution; the anisotropic noise case is treated in section \ref{sec:ISW-lensing}, using directly \planck\ simulations).
The maps are both analyzed in the full and partially masked cases, the mask being the \planck\ 2018 common mask ($\fsky=0.78$). Similarly to figure \ref{fig:loc-256}, figure \ref{fig:loc-2048} shows the averaged integrated bispectra determined from these maps and the corresponding standard deviations (from the Gaussian maps only) compared to the predictions from theory.
The observed integrated bispectra of the masked case are renormalized by the usual factor $\fsky^{-1}$ while all the theoretical lines correspond to the full sky case (integrated bispectrum and standard deviation).
Again, the observed results agree very well with the theory using the linear correction derived in this work (eq.\ \ref{eq:linear-correction}) to remove all the spurious non-Gaussian contribution due to the mask. 

Another illustration is provided in table \ref{tab:local} which contains the values of $\fnl$ determined for a full and a partially masked sky. For the latter, we give the results with and without the linear correction. Without this linear correction, the error bars are $\sim 50$ times bigger than in the full sky case. After applying it we recover the expected error bars with only a small increase.

\begin{table}
    \begin{center}
      \small
    \begin{tabular}{lccc}
    \hline
    & Unmasked & Masked (no linear correction) & Masked (with linear correction)\\
    \hline
    \multicolumn{1}{@{\hspace{0.5cm}}c@{\hspace{0.3cm}}}{$\fnlloc = 0$} & $1.25 \pm 0.67$ & $59 \pm 39$ & $1.39 \pm 0.73$\\
    \multicolumn{1}{@{\hspace{0.5cm}}c@{\hspace{0.3cm}}}{$\fnlloc = 20$} & $21.58 \pm 0.85$ & $79 \pm 39$ & $20.66 \pm 0.91$\\
    \hline
    \end{tabular}
    \end{center}
    \caption{Determination of the parameter $\fnlloc$ from Gaussian ($\fnlloc=0$) and non-Gaussian ($\fnlloc=20$) CMB simulations using the integrated bispectrum estimator (eq.\ \ref{eq:fnl}). The same 100 maps are analyzed in the full and partially masked (\planck\ 2018 common mask, $\fsky=0.78$) sky cases. Results are given with and without linear correction.}
    \label{tab:local}
\end{table}

The integrated bispectrum methodology developed in section \ref{sec:integrated-bispectrum} and \ref{sec:ib-estimator} passes successfully our tests with the local shape, showing a perfect agreement between theoretical predictions and the analyses of simulations with only slightly suboptimal results compared to a (full) bispectrum estimator.

\subsection{ISW-lensing}
\label{sec:ISW-lensing}

It is well-known that the gravitational lensing of CMB photons -- due to cosmic structures along the line-of-sight (affecting the small scales) -- and their shift in temperature as they go into wells and hills of the evolving late-time gravitational potentials -- known as the ISW effect (affecting the large scales) -- are correlated and therefore produce a large squeezed bispectrum \cite{Goldberg:1999xm}. This ISW-lensing shape has been successfully detected by the \planck\ collaboration at more than 3$\sigma$ \cite{Ade:2013ydc, Ade:2015ava, Akrami:2019izv}.

The ISW-lensing bispectrum template (see \cite{2011JCAP...03..018L} for details) is given by
\begin{equation}
    B_{\ell_1 \ell_2 \ell_3}^\mathrm{lens} = h_{\ell_1 \ell_2 \ell_3}^2 \left[ f_{\ell_1 \ell_2 \ell_3} C_{\ell_2}^{T\phi} C_{\ell_3}^{TT} + (5\text{ perm.}) \right]\,,
\end{equation}
where $C_{\ell}^{TT}$ is the lensed CMB power spectrum (temperature), $C_{\ell}^{T\phi}$ is the ISW-lensing cross power spectrum and the function $f_{\ell_1 \ell_2 \ell_3}$ is defined as
\begin{equation}
    f_{\ell_1 \ell_2 \ell_3} \equiv \frac{1}{2}\left[\ell_2(\ell_2 + 1) + \ell_3(\ell_3 + 1) - \ell_1(\ell_1 + 1)\right]\,.
\end{equation}

On the \planck\ legacy archive, there are several sets of lensed CMB simulations which are publicly available and they have been extensively tested at the non-Gaussian level already.%
\footnote{\url{https://pla.esac.esa.int}}
We use the CMB (temperature) simulations which have passed through the foreground-separation pipelines \texttt{SEVEM} \cite{Fern_ndez_Cobos_2012} and \texttt{SMICA} \cite{2008ISTSP...2..735C}. These maps have a Gaussian beam with a 5 arcmin FWHM. In addition to looking at the ISW-lensing integrated bispectrum in these simulations, we use them to determine the error bars and to compute the linear correction terms, necessary for the analysis of the \planck\ observational data provided in section \ref{sec:planck}. We also use the 300 corresponding simulated noise maps for the two cleaning techniques, from which we compute the noise power spectra.
After masking the maps with the \planck\ 2018 common mask ($\fsky = 0.78$), we apply the usual inpainting procedure, briefly described in section \ref{sec:ib_obs}. Then, the lensed primordial power spectrum for each component separation method is determined by averaging the power spectra extracted from the corresponding simulations. From these power spectra, we compute the covariance of the integrated bispectra using eq.\ \eqref{eq:ibisp_covariance}.

We use the $\tt{Plik}$ 2018 best fit cosmological parameters (see the first column of table 1 of \cite{Aghanim:2018eyx}) to determine the local bispectrum (recalling that however this shape is not very sensitive to the exact values of these parameters). We use the CAMB computer code to evaluate the temperature-lensing potential cross-power spectra $C_{\ell}^{T\phi}$, necessary to compute ISW-lensing bispectra. Then, we derive, for both bispectrum shapes, their respective integrated bispectra using eq.\ \eqref{eq:ibisp}. We keep using the simple step function patches ($w_\ell=1$ for $\ell\leq10$ and 0 otherwise) introduced earlier.

Because of the anisotropic noise and the mask, the linear correction is necessary. We apply the procedure developed in section \ref{sec:linear-correction} to the CMB simulations (including the noise simulations).

\begin{figure}
  \centering
  \includegraphics[width=0.49\linewidth]{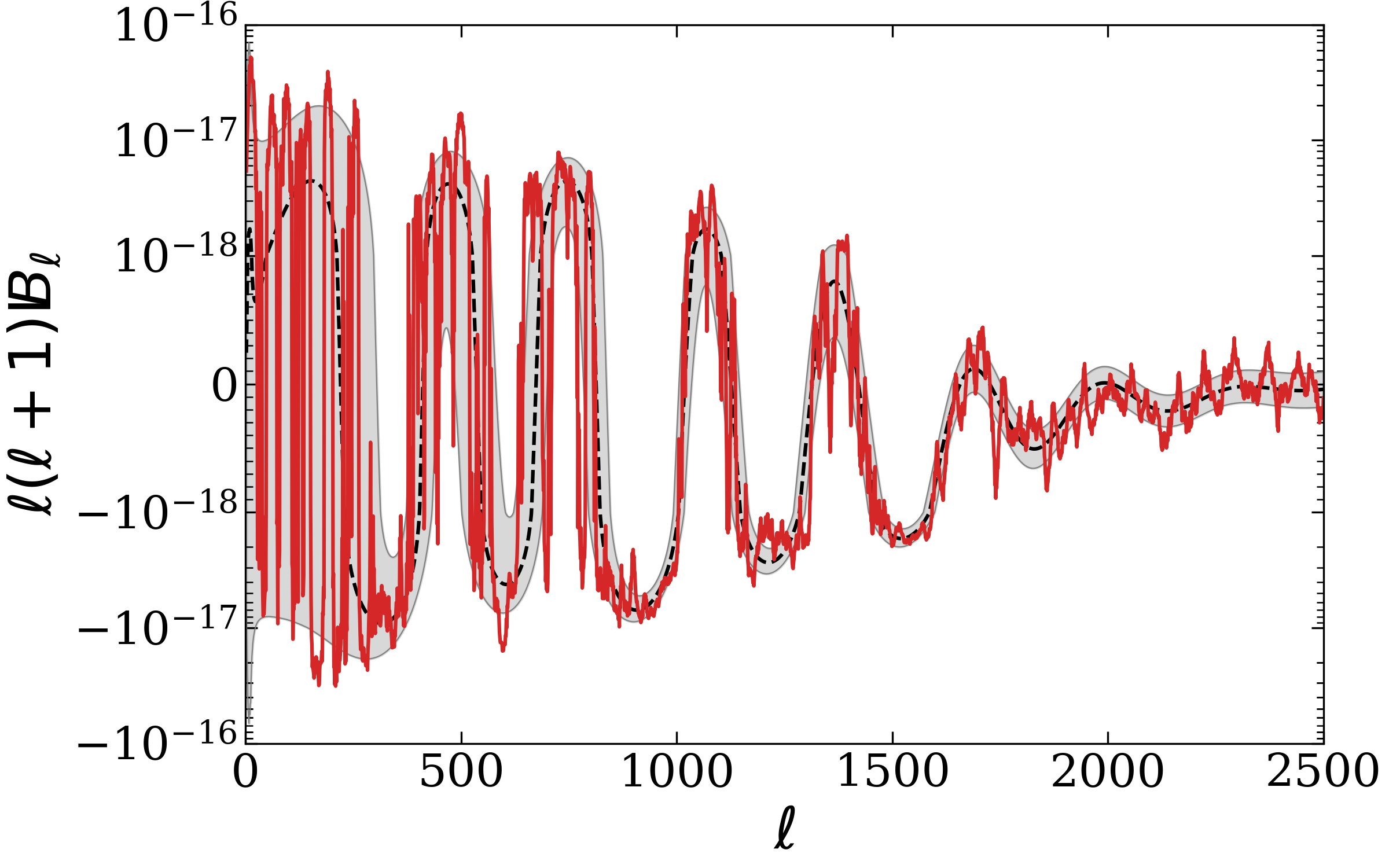}
  \hfill
  \includegraphics[width=0.49\linewidth]{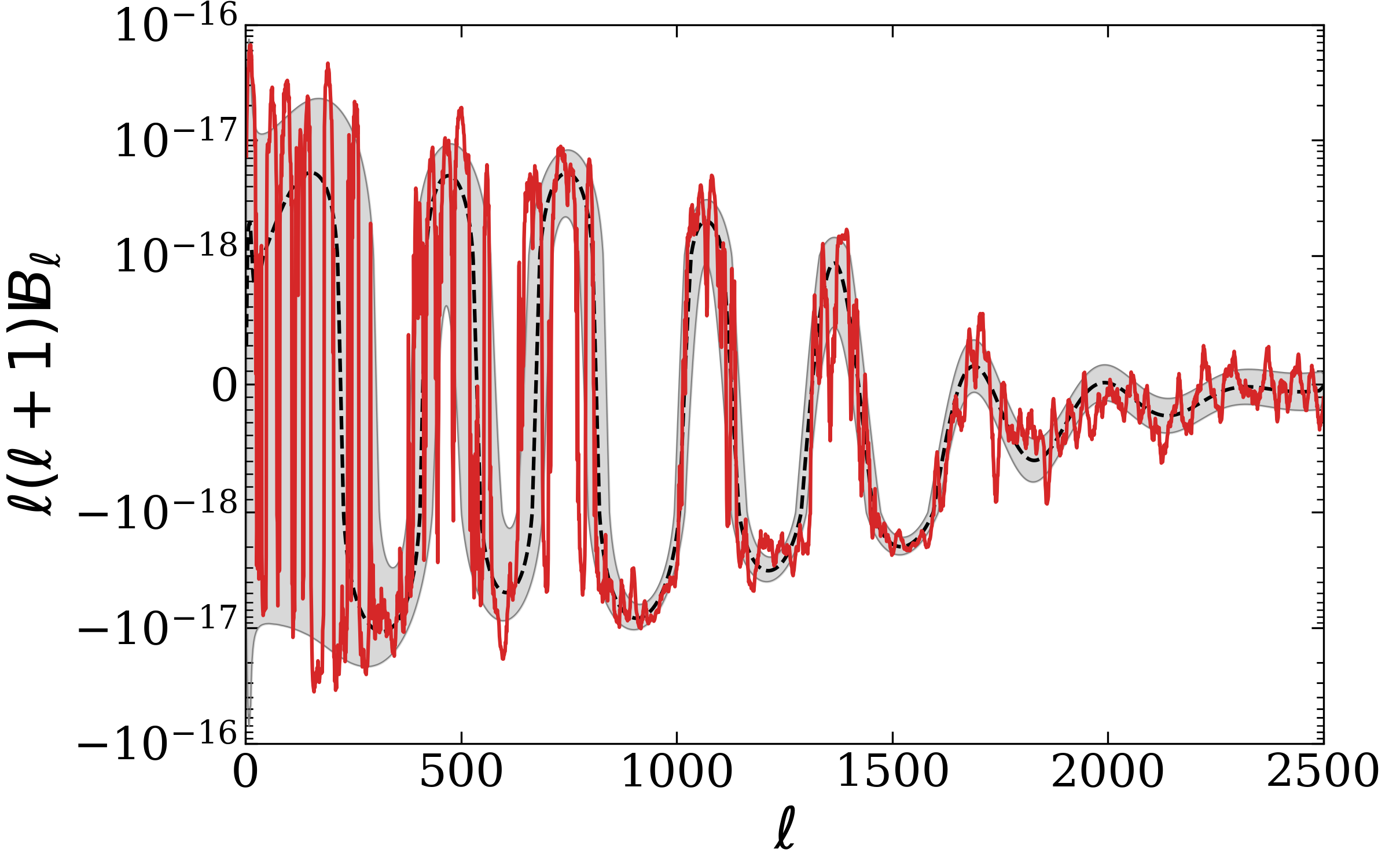}
  \caption{The integrated bispectra from \planck\ 300 CMB simulations (solid red line, \texttt{SEVEM} on the left and \texttt{SMICA} on the right). The black dashed lines correspond to the theoretical ISW-lensing integrated bispectrum computed using eq.\ \eqref{eq:ibisp} with its expected standard error (68\% CL) in grey, computed with eq.\ \eqref{eq:ibisp_covariance}. Note that all quantities are multiplied by the factor $\ell(\ell+1)$ for visibility. Moreover, the scale is linear for values in $[-10^{-18}, 10^{-18}]$ and logarithmic otherwise.}
  \label{fig:planck-simulations}  
\end{figure}

In figure \ref{fig:planck-simulations} we show the averaged integrated bispectra of the \texttt{SEVEM} and \texttt{SMICA} simulations, which as expected correspond to the theoretical ISW-lensing shape. In table \ref{tab:planck-simulations} we indicate the averaged $\fnl$ of these maps with their standard errors.
As usual, the presence of the ISW-lensing bispectrum induces a non-negligible bias on $\fnlloc$, if the two shapes are studied independently with the simple estimator (eq.\ \ref{eq:fnl}). The ISW-lensing bias on the $\fnlloc$ integrated bispectrum estimate   is  compiuted to be 4.3, which is smaller than the value around 7 found with an optimal bispectrum estimator (see table 3 of \cite{Akrami:2019izv}). The reason is that the ISW-lensing bispectrum is less peaked around the squeezed limit than the local shape, thus its integrated bispectrum does not contain the information of some important multipole triplets (an illustration at the level of the bispectrum is provided later in figure \ref{fig:weights}). This effect can also be seen in the error bar on $\fnl$, which, for this template,  becomes three times larger with respect to the optimal, full bispectrum analysis. By comparison, the error for our $\fnlloc$ estimate grows only by 25\%, compared to the full bispectrum case.

Besides directly subtracting the expected ISW-lensing bias on $\fnlloc$, we also perform a joint analysis of the two templates, using the estimator (eq.\ \ref{eq:fnl-joint}). 
This yields the expected result: we measure the ISW-lensing result at the predicted level, while $\fnlloc$ is consistent with $0$. The joint analysis also produces a small but non negligible increase of the error bars (less than 10\%). This is larger than the increase of error bars for joint analysis with optimal bispectrum estimators. In the full bispectrum analysis, the local and ISW-lensing shapes are in fact not enough correlated to have a visible effect on the error bars. The correlation coefficient between the two shapes, defined in eq.\ \eqref{eq:correlation-coefficient}, is 0.53 in the integrated bispectrum case, while a similar computation at the bispectrum level gives a result slightly smaller than 0.3. 

As for the local shape in the previous section, the integrated bispectrum passes successfully comparison test between theoretical predictions and measurements from robust simulations. This method turns out to be however well-suited to look for the ISW-lensing shape in data, as the results are clearly suboptimal.

\begin{table}
  \begin{center}
    \small
  \begin{tabular}{lcc}
  \hline
  & Local  & ISW-lensing \\
  \hline
  \hfil \texttt{SMICA} & & \\
  Independent & $4.59 \pm 0.43~(7.5)$ & $1.02 \pm 0.06~(1.0)$ \\
  Joint & $0.33 \pm 0.49~(8.5)$ & $1.00 \pm 0.07~(1.2)$  \\
  \hfil \texttt{SEVEM} & & \\
  Independent & $4.50 \pm 0.45~(7.7)$ &  $1.02 \pm 0.06~(1.0)$ \\
  Joint & $0.33 \pm 0.50~(8.6)$  & $1.00 \pm 0.07~(1.2)$ \\
  \hline
  \end{tabular}
  \end{center}
  \caption{Determination of the $\fnl$ parameters of the local and ISW-lensing shapes from \planck\ CMB simulations using the integrated bispectrum estimator (eq.\ \ref{eq:fnl}). Results are given with the standard errors (68\% CL on the average of 300 maps) and the corresponding standard deviations inside parentheses.}
  \label{tab:planck-simulations}
\end{table}

\subsection{ISW-tSZ-tSZ}
\label{sec:ISW-tSZ-tSZ}
The tSZ effect, generated by hot electrons that belong, for the most part, to galaxy clusters sitting in potential wells, strongly correlates with the ISW effect \cite{HernandezMonteagudo:2004xg}.
In \cite{Hill:2018ypf} it has been discussed how a specific three-point cross-correlator, ISW-tSZ-tSZ, is particularly important, and might constitute a bright foreground for local type non-Gaussianity searches.
While it goes beyond the scope of this paper --- more focused in this specific correlator as a potential source of bias on $\fnlloc$ --- we clarify that
a complete analysis aimed at detecting ISW-tSZ-tSZ must be multi-frequency and account for component separation.

The tSZ effect generates a spectral distortion that, at first order, has a distinct Compton-$y$ shape
(see, e.g., \cite{Chluba:2012dv,Remazeilles:2018laq} for refinements and considerations on relativistic corrections that we will not use here for simplicity).%
\footnote{Accounting for relativistic corrections would anyway suppress the signal, strengthening out claims we make later in our analysis.}
Therefore, temperature fluctuations at a given frequency are given by
%-------------------------
\begin{equation}
    \frac{\Delta T}{T} (\nu) = g(\nu) \, y,
    \quad g(\nu)
    \equiv
    x \coth (x/2) - 4 \, ,
\label{eq:y-distortion}
\end{equation}
%-------------------------
where $x \equiv h \nu/(k_\text{B} T)$ and $y$ is an amplitude parameter that we predict using a halo-model approach.

We closely follow the methodology and notation of \cite{Hill:2018ypf}.
The angle-averaged bispectrum of ISW-tSZ-tSZ is given by
\begin{equation}
\begin{split}
    B_{\ell_1 \ell_2 \ell_3}
    =
    & h_{\ell_1 \ell_2 \ell_3}^2 \,
    g(\nu_2) \, g(\nu_3)
    \int \diff z \, \mathcal{I}_{\ell_1}(z)
    \left[ 
        P_\text{lin} \bigg( \! \frac{\ell_1 + \frac{1}{2}}{\chi(z)} \! \bigg) \!
        \int \diff M \td{n}{M} \, b(M,z) \, \tilde{y}_{\ell_2}(M,z) \, \tilde{y}_{\ell_3}(M,z) \, +
    \right.
\\
    & +
    P_\text{lin} \bigg( \! \frac{\ell_2 + \frac{1}{2}}{\chi(z)} \! \bigg) \!
    \int \diff M \td{n}{M} \, b(M,z) \, \tilde{y}_{\ell_2}(M,z)
    \int \diff M' \td{n}{M'} \, b(M',z) \, \tilde{y}_{\ell_3}(M',z) \frac{M'}{\rho_m} \, +
\\
    & +
    \left.
        P_\text{lin} \bigg( \! \frac{\ell_3 + \frac{1}{2}}{\chi(z)} \! \bigg) \!
        \int \diff M \td{n}{M} \, b(M,z) \, \tilde{y}_{\ell_3}(M,z)
        \int \diff M' \td{n}{M'} \, b(M',z) \, \tilde{y}_{\ell_2}(M',z) \frac{M'}{\rho_m}
    \right] \,
\\
    & + \text{2 cyclic permutations} \, ,
\end{split}
\label{eq:Tyy-template}
\end{equation}
%-------------------------
where
%-------------------------
\begin{equation*}
    \mathcal{I}_\ell (z)
    =
    \frac{3 \, \Omega_m \, H_0^2}{c^2 \big(\ell + \frac{1}{2}\big)^2} \,
    \chi^2(z) \, D(z) \td{}{z}
    \left(
        \frac{D(z)}{a(z)}
    \right) .
\end{equation*}
%-------------------------
The term shown explicitly refers to a configuration in which $\ell_1$ is associated to an ISW fluctuation, and $\ell_2$ and $\ell_3$ to tSZ fluctuations.
Therefore, the permutations affect the three $\ell_i$ and the frequencies $\nu_i$.

In eq.\ \eqref{eq:Tyy-template} we introduced the halo mass function (HMF) $\td{n}{M}$, halo bias $b$, and the harmonic transform of the halo Compton-y profile
%-------------------------
\begin{equation}
    \tilde{y}_\ell
    =
    4 \pi \frac{\sigma_\text{T}}{m_\text{e} c^2}
    \frac{r_s}{\ell_s^2}
    \int \diff x \, x^2 \, j_0\Big(\frac{\ell}{\ell_s} x \Big) P_\text{e} (x, M, z) \, .
\label{eq:ytilde}
\end{equation}
%-------------------------
Here $\sigma_\text{T}$ is the \Thomson cross section, $m_\text{e}$ is the electron rest mass and $j_0$ is the spherical Bessel function of order 0.
$P_\text{e}$ is the electron pressure profile, parametrized as described below, and $r_s$ and $\ell_s \equiv d_A/r_s$ are respectively the characteristic scale of the electron pressure profile and the associate multipole moment, $d_A$ being the angular diameter distance.

To calculate the ISW-tSZ-tSZ bispectrum we employ two different sets of parametrizations for the fitting functions that appear in eq.\ \eqref{eq:Tyy-template}.

In the first case (H18), following \cite{Hill:2018ypf, Komatsu:2002wc},  we integrate over the virial masses , and use the concentration mass relation provided in \cite{Duffy:2008pz} to convert the virial masses into the overdensity masses needed in the HMF and pressure profile.
The HMF is the one from \cite{Tinker:2008ff}, with the updated parameters and bias from \cite{2010ApJ...724..878T}, with the parameters calibrated on $M_{200,m}$.
In this case, we use the pressure profile from \cite{Battaglia:2011cq}.
Consequently, the scale radius in eq.\ \eqref{eq:ytilde} is $r_s = r_{200,c}$.

In the second case (Pl18) we integrate directly over the overdensity mass $M_{500,c}$, as in, e.g., \cite{Aghanim:2015eva, Bolliet:2017lha}, so that a conversion to the virial masses is never required.
Again, we use the same HMF as before, but convert $M_{500,c}$ into $M_{\Delta,m}$, and fit the parameters of their table 4 to the appropriate value of $\Delta = 500 \rho_c / \rho_m$.
Furthermore we use the approximation $\diff \ln M_{\Delta,c} / \diff \ln M_\text{vir} \approx 1$.
Finally, we use the pressure profile from \cite{Arnaud:2009tt} (therefore $r_s = r_{500,c}$) and we also consider an hydrostatic mass bias rescaling the true mass of the halo, i.e., we use $P_\text{e} (x, \tilde{M}, z)$ where $\tilde{M} = (1-b_H)M$. We fix $b_H=0.2$ \cite{Aghanim:2015eva} throughout the analysis.

The two cases also differ for the choice of cosmological parameters.
In H18 we use the best-fit cosmology of \cite{Komatsu:2008hk} as in \cite{Hill:2018ypf}
(%
$h_0 = 0.702$,
$\sigma_8 = 0.817$,
$n_s = 0.962$,
$\Omega_m = 0.277$,
$\Omega_b = 0.0459$,
$\tau = 0.088$%
),
whereas in Pl18 we use the updated cosmological parameters from \cite{Aghanim:2018eyx} 
(%
$h_0 = 67.32$,
$\sigma_8 = 0.812$,
$n_s = 0.96605$,
$\Omega_m = 0.3158$,
$\Omega_b = 0.04939$,
$\tau = 0.0543$%
).
Notice that in \cite{Aghanim:2015eva} it has been shown that the \planck\ SZ measurement prefer lower $\Omega_m$ and $\sigma_8$ than the one used here, if we assume as we do that $b_H=0.2$.
However for simplicity we stick to the primary CMB best-fit parameters used in the rest of the paper, with the understanding that the amplitude of the ISW-tSZ-tSZ bispectrum would probably be overestimated.

\begin{figure}
    \centering
    \includegraphics[width=0.66\linewidth]{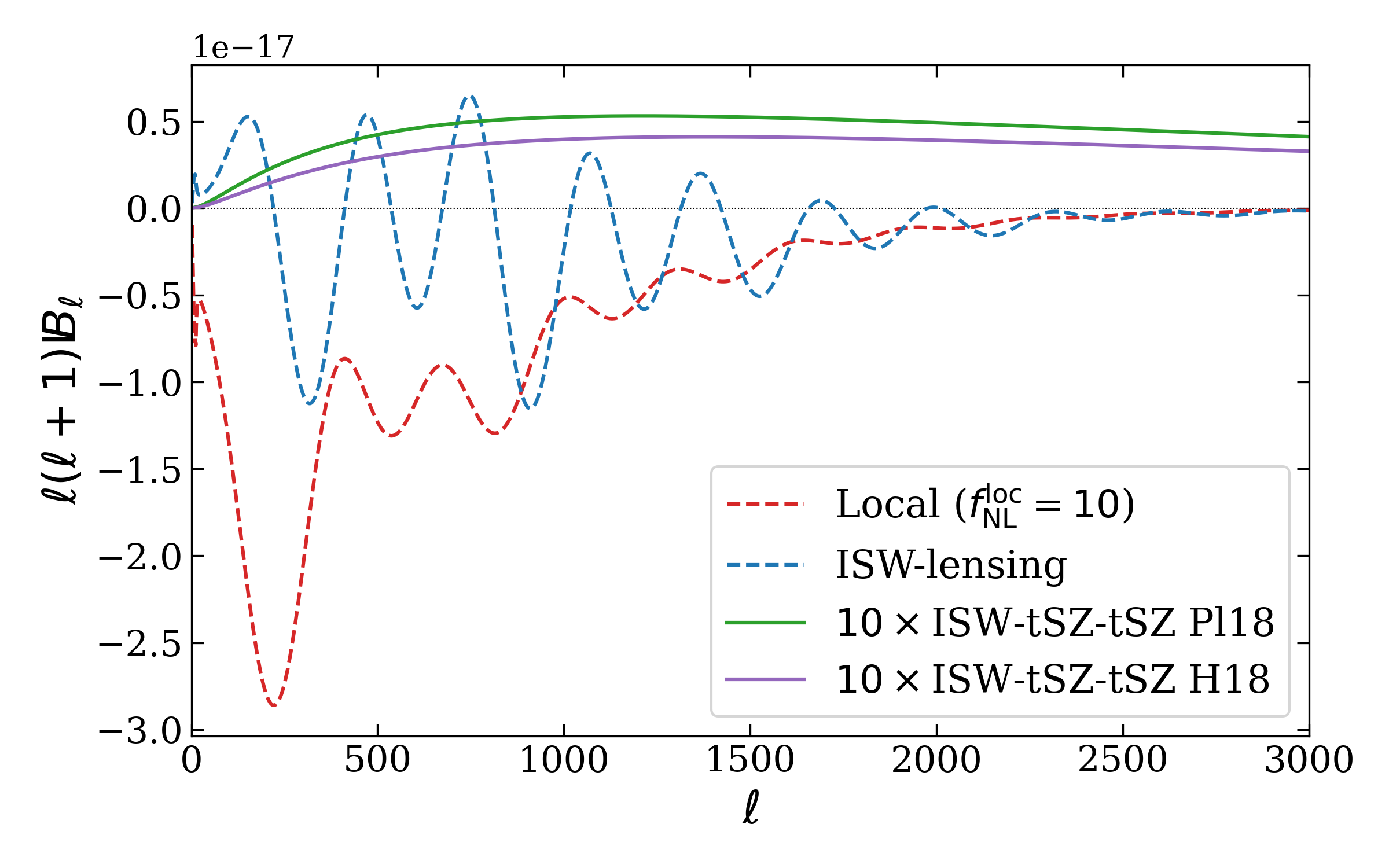}
    \caption{The integrated bispectrum of the ISW-tSZ-tSZ shape at 143 GHz, computed from eqs.\ \eqref{eq:Tyy-template} and \eqref{eq:ibisp} for the two parametrizations H18 (solid purple line) and Pl18 (solid green line). For comparison, the local and ISW-lensing shapes (Pl18 cosmological parameters) are also shown (red and blue dashed lines respectively). All these patches have been computed using step function multipoles patches ($w_\ell = 1$ for $\ell\leq10$ and 0 otherwise). Note that the local shape is represented for $\fnlloc=10$ and the amplitude of the ISW-tSZ-tSZ shapes has been multiplied by 10.}
    \label{fig:Tyy-ibisp}  
\end{figure}

\begin{figure}
    \centering
    \includegraphics[width=0.8\textwidth, clip]{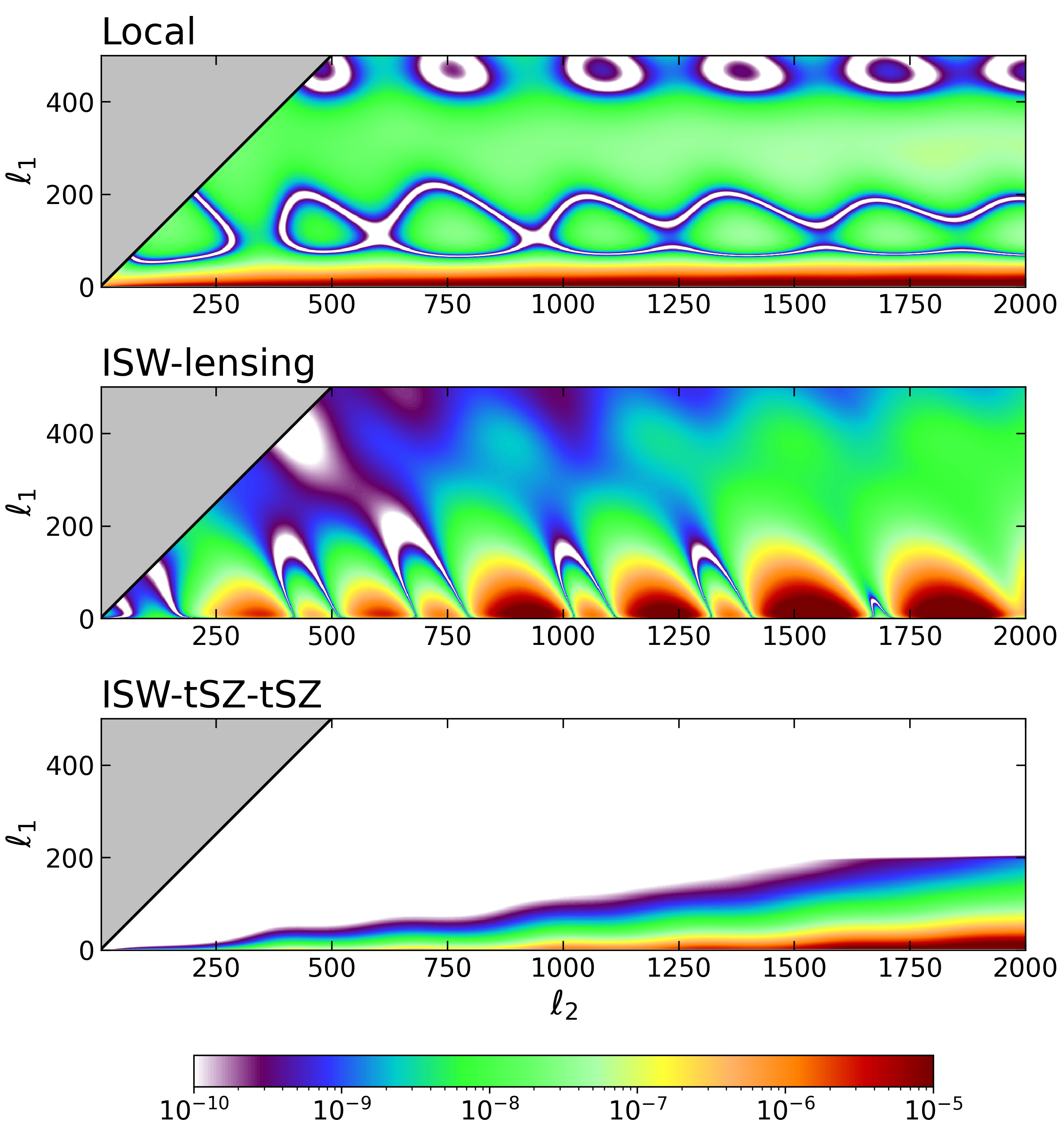}
    \caption{The bispectral weights of three bispectra peaking in the squeezed limit (local, ISW-lensing, ISW-tSZ-tSZ) computed using eq.\ \eqref{eq:bispectrum-estimator}. They are represented as a function of $\ell_1$ and $\ell_2$ and summed over $\ell_3$, showing only $\ell_1\leq500$ to focus on the squeezed limit. These weights are normalized (the sum of all the weights is 1), hence the frequency factors $g(\nu)$ of the ISW-tSZ-tSZ shape \eqref{eq:Tyy-template} cancel out. Moreover, in the normalization factor, the bispectrum variance is computed in a noiseless case. Note also that the colour scale is logarithmic.}
    \label{fig:weights}  
\end{figure}

From the ISW-tSZ-tSZ bispectrum shape given in eq.\ \eqref{eq:Tyy-template}, it is straightforward to compute the corresponding integrated bispectrum using eq.\ \eqref{eq:ibisp}. The results, determined using the same step function patches ($w_\ell = 1$ for $\ell\leq10$ and 0 otherwise) as in sections \ref{sec:local} and \ref{sec:ISW-lensing} and for both parametrizations discussed above (H18 and Pl18), are shown in figure \ref{fig:Tyy-ibisp}. Note that these shapes are computed at a frequency of 143 GHz, and that only the amplitude changes with the frequency (it is 2.1 times larger at 100 GHz, 4.6 times larger at 353 GHz). From the plot, it is clear that the ISW-tSZ-tSZ shape is small compared to the ISW-lensing template in the multipole range probed by \planck. However for smaller scales ($\ell$ above 2000), the situation changes and it becomes dominant.
It is also interesting to have a more complete view of this template as shown in figure \ref{fig:weights} where we compare the angle-averaged bispectra of the local, ISW-lensing and ISW-tSZ-tSZ shapes (to be more precise, these are bispectral weights in multipole space which are properly defined in appendix \ref{sec:weights}). They are computed in the case of an ideal experiment, using the \planck\ 2018 best fit cosmology. Shapes which peak in the squeezed configurations have the largest weights along the bottom line (i.e.\ small $\ell_1$).
As expected, for the three bispectra shown here, this is the behaviour we can observe. Indeed in the plots the red zones correspond to the important configurations (and because the scale is logarithmic most of the rest is actually negligible). For the local and ISW-tSZ-tSZ templates, the choice of considering only the smaller multipoles ($\ell\leq\ell_w=10$ with the step function patches) for the integrated bispectrum is justified as most of the non-Gaussian information will be included. However, in the case of ISW-lensing the weights are also large for less squeezed configurations (with $\ell_1$ up to 50 or even more). This is in agreement with the poorer performance of the integrated bispectrum we noticed in the previous section for this shape specifically.
While this is beyond the scope of this paper, we point out that the smooth behaviour of the ISW-tSZ-tSZ shape makes it suitable to study with binned or modal bispectrum estimators.

\subsection{\planck\ temperature analysis}
\label{sec:planck}

Using the integrated bispectrum estimator, we perform a new analysis of non-Gaussianity in  \planck\ temperature data (2018 release), considering the shapes described in the previous three sections. For the latest official results using several standard bispectrum estimators, see \cite{Akrami:2019izv}.

As mentioned earlier, we use the \texttt{SEVEM} and \texttt{SMICA} CMB temperature maps and their corresponding simulations which are available on the \planck\ Legacy Archive. We apply the pipeline described in \ref{sec:implementation} to these maps (see section \ref{sec:ISW-lensing} for all the details). In table \ref{tab:planck}, we report $\fnl$ values from the observed maps, using the error bars and the linear corrections computed from the simulations. These results are fully compatible with the \planck\ non-Gaussianity analysis conclusions, there is no detection of primordial non-Gaussianity. For completeness, we also apply this method to two other standard primordial templates, the equilateral and orthogonal shapes. However, as they do not peak in the squeezed configuration, the results are strongly suboptimal and we do not include them in detail here ($\sigma(\fnl^\mathrm{equil})\approx1300$ and $\sigma(\fnl^\mathrm{ortho})\approx230$).

\begin{table}
    \begin{center}
      \small
      \begin{tabular}{lcccc}
        \hline
        & Local & ISW-lensing \\
        \hline
        \hfil \texttt{SMICA} & & \\
        Independent & $7.1 \pm 7.5$  & $0.8 \pm 1.0$ \\
        Joint & $5.0 \pm 8.5$ &  $0.5 \pm 1.2$    \\
        \hfil \texttt{SEVEM} & &\\
        Independent & $11.1 \pm 7.7$ & $1.2 \pm 1.0$ \\
        Joint & $8.6 \pm 8.6$ & $0.6 \pm 1.2$  &   \\
        \hline
        \end{tabular}
      \end{center}
    \caption{Determination of the $\fnl$ parameters of the local, equilateral, orthogonal and ISW-lensing shapes from the \texttt{SMICA} and \texttt{SEVEM} foreground-cleaned maps using the integrated bispectrum estimator (eq.\ \ref{eq:fnl}).}
    \label{tab:planck}
\end{table}

To study the ISW-tSZ-tSZ shape, we also use the foreground-cleaned single-frequency maps at 100 and 143 GHz provided by \texttt{SEVEM} to which we apply the exact same methodology as above. Note that these maps have different levels of noise and lower resolutions (their FWHM are 9.66 and 7.27 arcmin at 100 and 143 GHz respectively), thus we use lower cuts  ($\lmax=$1500 and 2000 at 100 and 143 GHz respectively).

The results for $\fnl$ of the three squeezed shapes discussed in this paper can be found in table \ref{tab:planck-channels}. The results for the local shape are compatible with $\fnlloc = 0$ as expected, with larger error bars than in table \ref{tab:planck}. Concerning the ISW-tSZ-tSZ template, its $\fnl$ parameter is normalized in such a way that its expected value is 1 at each frequency for the \planck\ 2018 best fit cosmology. Here, the error bars are one order of magnitude too large in light of $\fnl=1$, which eliminates the possibility of a detection in a single-frequency channel. This also means that the bias produced by the ISW-tSZ-tSZ bispectrum on the local shape is negligible here. Hence this possible contamination of primordial non-Gaussianity does not have an influence on the \planck\ results, even if it should be taken carefully into consideration in future experiments going to higher $\lmax$ (more on that in section \ref{sec:forecasts}).
Even though the ISW-tSZ-tSZ integrated bispectrum is $\sim2.1$ times larger at 100 GHz than at 143 GHz and thus the corresponding error bar on its $\fnl$ should be $\sim2.1$ smaller (because it scales like $g(\nu)^{-2}$, see section \ref{sec:ib-estimator} for details) if the experimental conditions (noise and beam) were the same. This is not the case here because of the lower resolution (and lower $\lmax$), as it actually even makes it slightly harder to detect in the low frequency channel. In table \ref{tab:planck-channels}, we also indicate the expected error bars of an optimal full bispectrum estimator in the independent case to show that the integrated bispectrum performs almost as well on the ISW-tSZ-tSZ shape as for the local shape in comparison to these estimators. Moreover, this shows that even with these optimal estimators (and the larger amount of computations they require), the influence of ISW-tSZ-tSZ cannot be detected by \planck\ in these frequency channels. We also report the results of the joint analysis of the three squeezed shapes, where the error bars or the local are around two times bigger. The main reason is that here the local and ISW-tSZ-tSZ are very correlated (with a coefficient of -0.87 and -0.81 at 100 and 143 GHz respectively), an issue already discussed in \cite{Hill:2018ypf}. However, due to the negligible expected bias produced by ISW-tSZ-tSZ on the other amplitude parameters, at \planck\ resolution, this joint analysis is not necessary when testing for the primordial local signal.

\begin{table}
    \begin{center}
      \small
      \begin{tabular}{lccc}
        \hline
        & Local & ISW-lensing & ISW-tSZ-tSZ  \\
        \hline
        \hfil \bf{100 GHz} & & & \\
        Independent & $11.8 \pm 10.9~(8.8)$ & $0.4 \pm 1.9$ & $-5.8 \pm 8.0~(5.5) $   \\
        Joint (2 shapes) & $12.3 \pm 11.2$ & $-0.2\pm 1.9 $  & \\
        Joint (3 shapes) & $19.5 \pm 23.2$ & $0.1 \pm 1.9$  & $6.1 \pm 16.9$\\
        \hfil \bf{143 GHz} & & & \\
        Independent & $10.3 \pm 8.7~(6.4)$ & $0.7 \pm 1.2$ & $-4.1 \pm 7.5~(4.4)$   \\
        Joint (2 shapes) & $10.2 \pm 9.4$ & $0.0\pm 1.3$  & \\
        Joint (3 shapes) & $19.0 \pm 14.1$ & $0.3 \pm 1.3$  & $9.9 \pm 12.2$\\
        \hline
        \end{tabular}
      \end{center}
    \caption{Determination of the $\fnl$ parameters of the local, ISW-lensing and ISW-tSZ-tSZ shapes from the \texttt{SEVEM} foreground-cleaned maps at 100 and 143 GHz using the integrated bispectrum estimator (eq.\ \ref{eq:fnl}). Inside the parentheses, we give the Fisher error bars from an optimal estimator using the full bispectrum for comparison.}
    \label{tab:planck-channels}
  \end{table}

\begin{figure}
  \centering
  \includegraphics[width=0.49\linewidth]{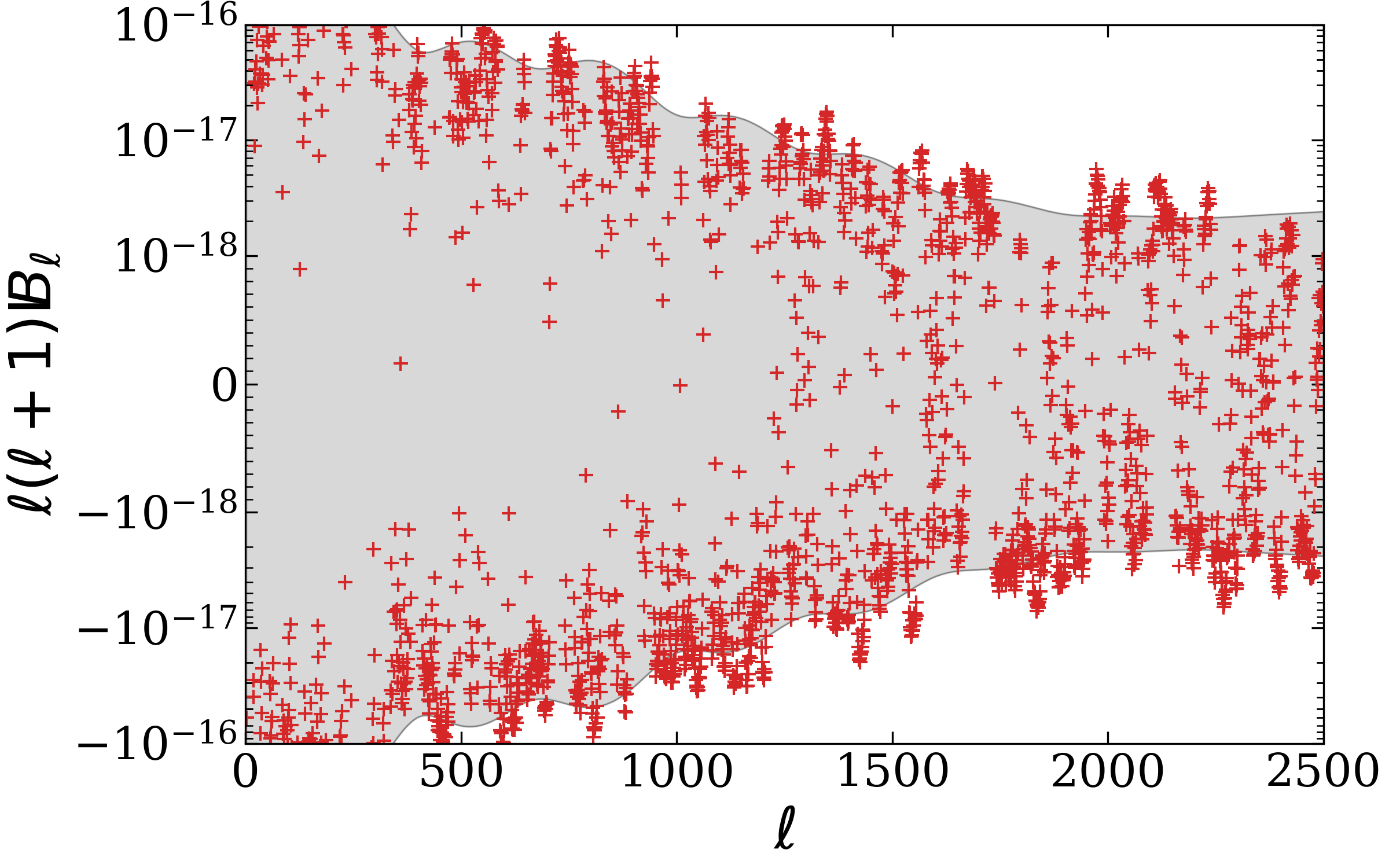}
  \hfill
  \includegraphics[width=0.49\linewidth]{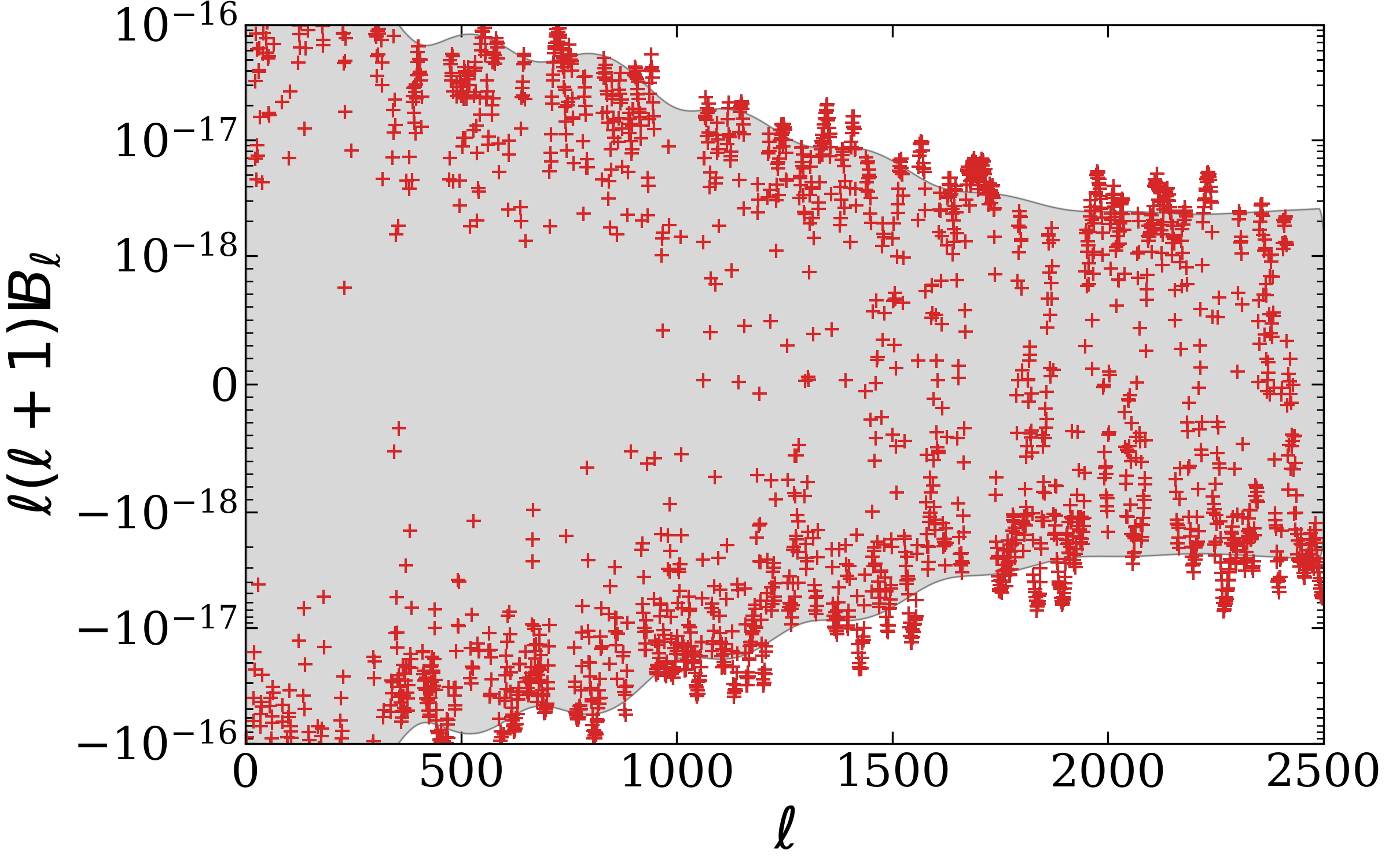}
  \includegraphics[width=0.49\linewidth]{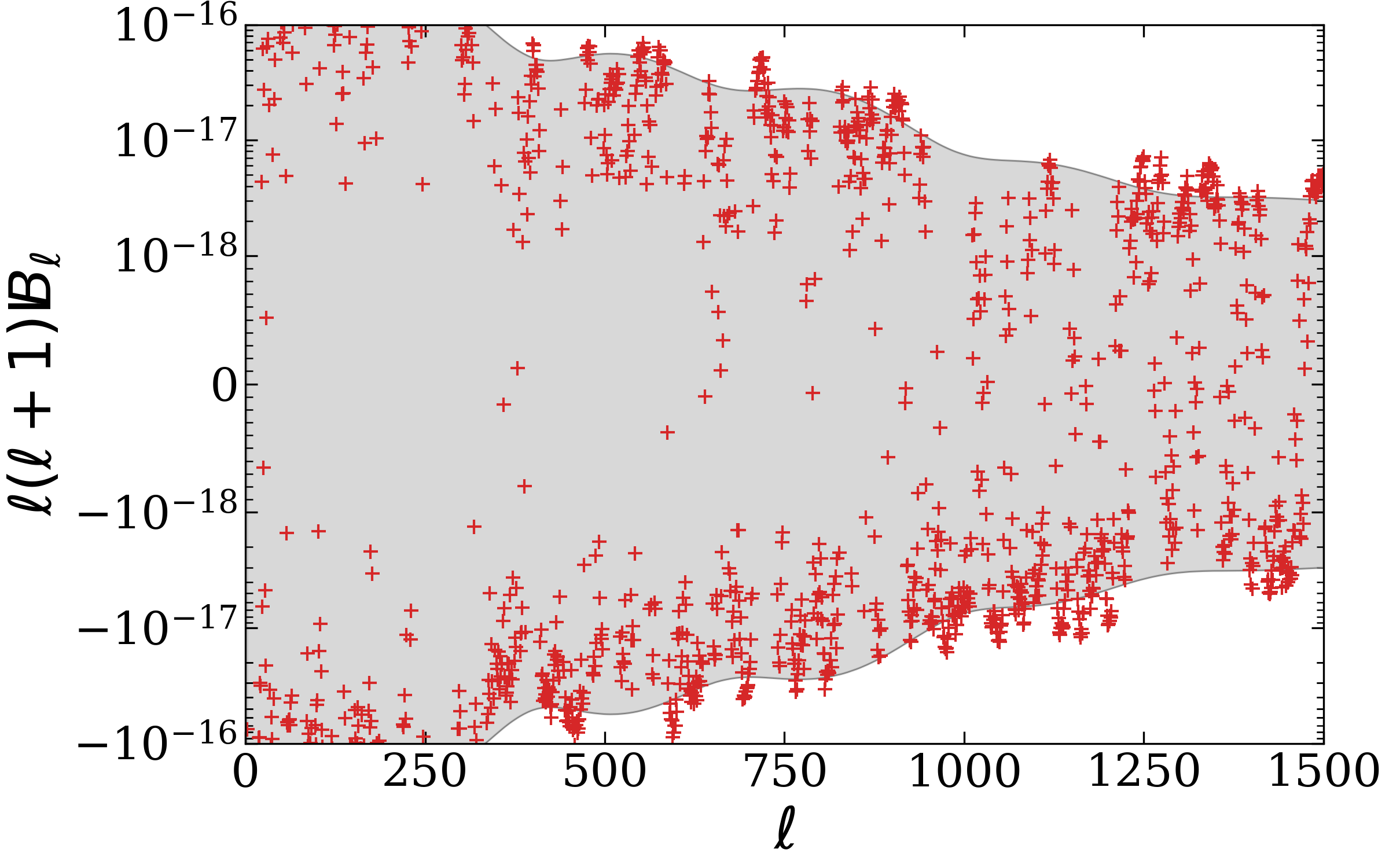}
  \hfill
  \includegraphics[width=0.49\linewidth]{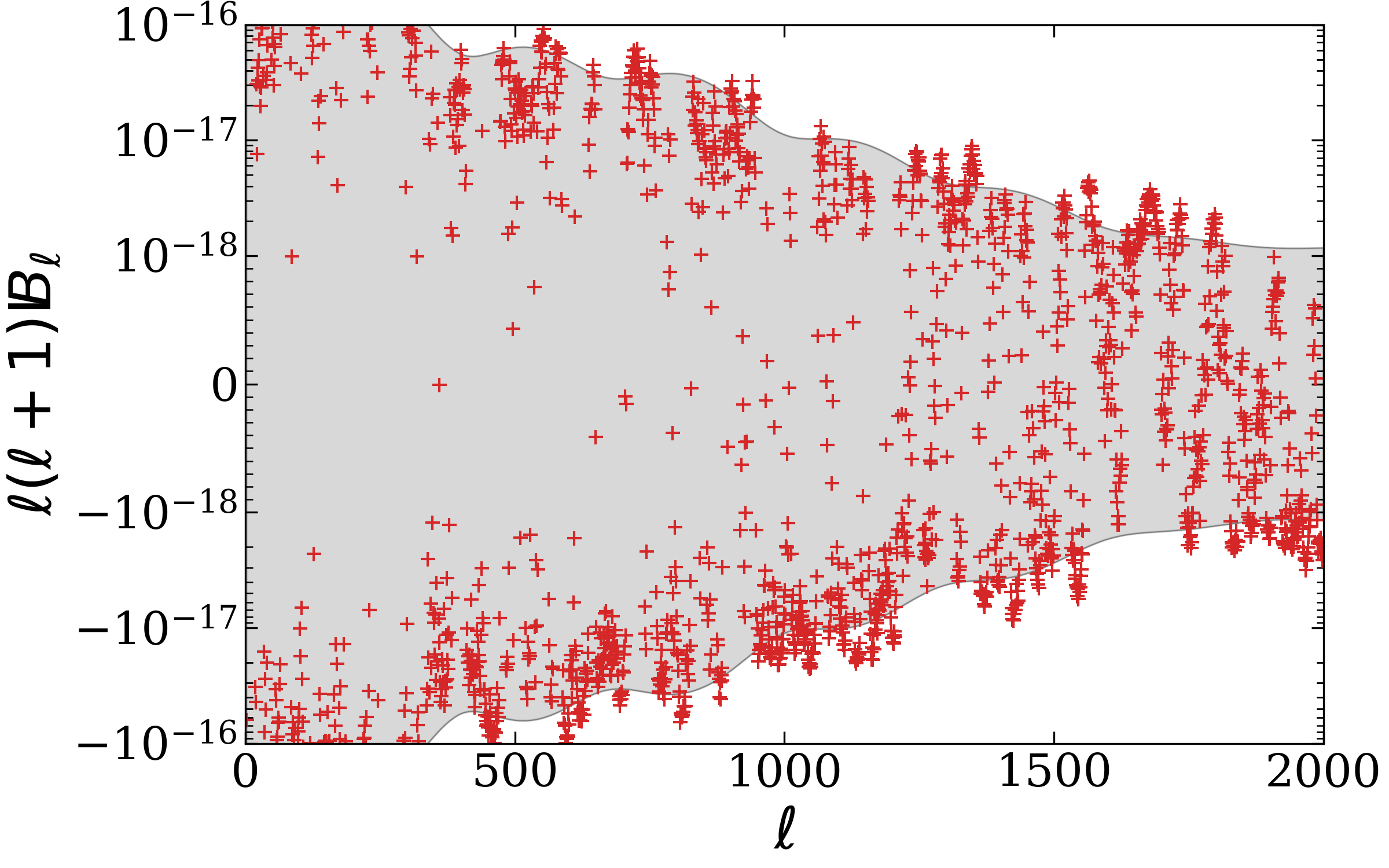}
  \caption{The observed integrated bispectrum from \planck\ CMB temperature data. The top pannels correspond to the \texttt{SEVEM} (left) and \texttt{SMICA} (right) maps, where the measured integrated bispectrum is in red and the grey areas indicate the expected error bar (68\% CL) around zero. The bottom pannels are the same for the single-frequency maps of \texttt{SEVEM} (100 GHz on the left and 143 GHz on the right). Note that the scale is linear for values in $[-10^{-18}, 10^{-18}]$ and logarithmic otherwise. Moreover, the top and bottom pannels have different $\lmax$.}
  \label{fig:planck-observations}  
\end{figure}

In addition to search for specific shapes in the data, we can also look directly at the observed data itself in figure \ref{fig:planck-observations}. For the four maps studied in this section (\texttt{SEVEM}, \texttt{SMICA}, \texttt{SEVEM} 100 GHz and \texttt{SEVEM} 143 GHz), the corresponding integrated bispectra show no sign of deviation from zero.

\subsection{Thermal dust}
\label{sec:dust}

Small grains of dust present of the interstellar medium of our galaxy absorb the starlight in the UV range and re-emit it as a thermal radiation at millimeter wavelengths. This emission is the brightest and has larger brightness fluctuations (i.e. small-scale fluctuations) in some regions of the sky corresponding to the largest clouds of dust (i.e. large-scale fluctuations) because they are also the thickest along the line of sight (see \cite{MivilleDeschenes:2007ya} for a discussion on the statistical properties of the dust). The resulting correlations between small and large-scale modes produce a large squeezed bispectrum. 

In non-Gaussianity analysis, this foreground is usually treated as a contamination of the primordial signal and removed from the maps using one of the numerous cleaning techniques available, one can also study its bispectrum as done recently in \cite{Jung:2018rgf,Coulton:2019bnz,Akrami:2019izv} with a binned bispectrum approach.

\begin{figure}
    \centering
    \includegraphics[width=0.66\textwidth]{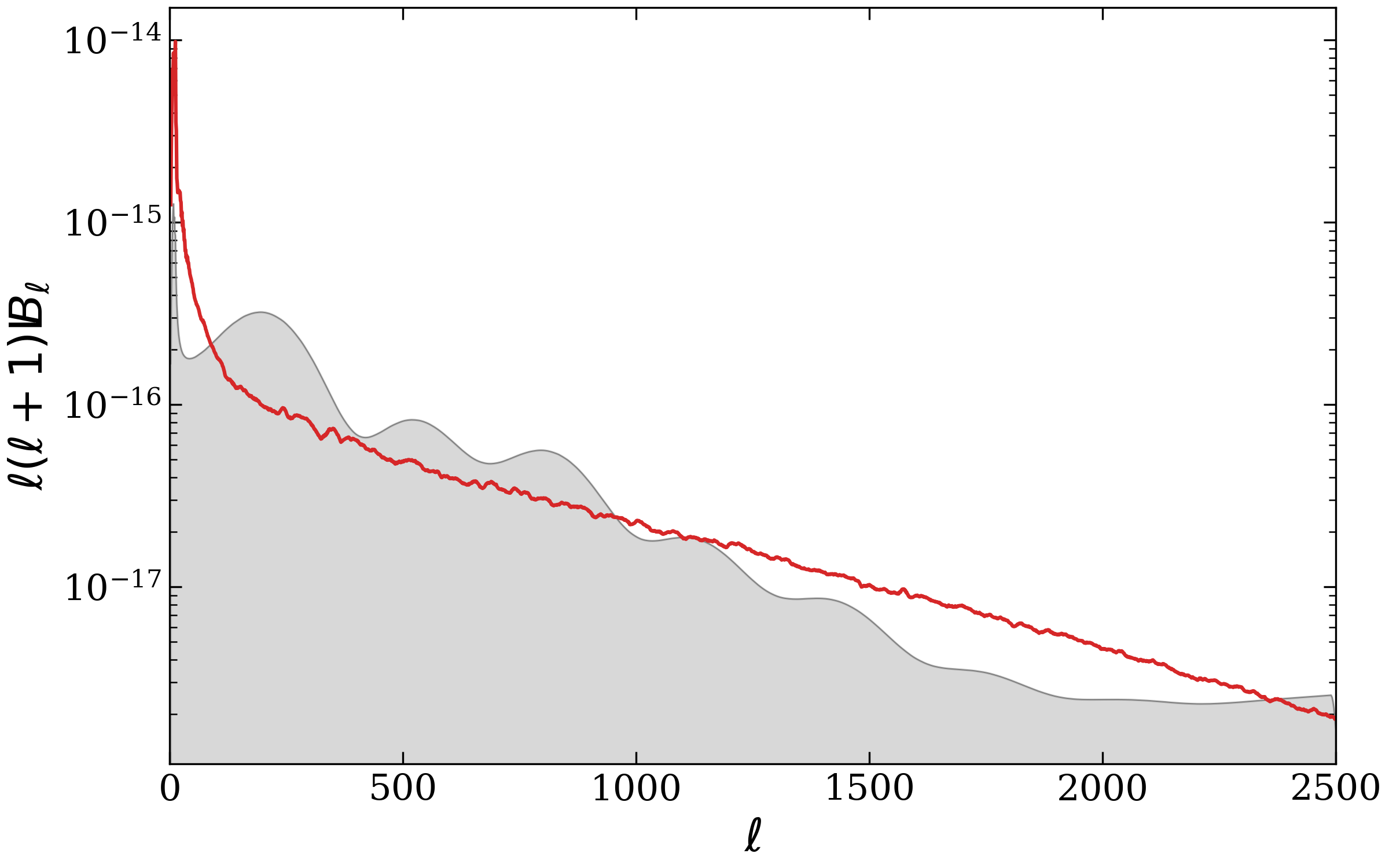}
    \caption{The integrated bispectrum of the thermal dust emission. In red, the integrated bispectrum determined from the dust \planck\ \texttt{Commander} map at 143 GHz. In grey, the standard deviation (68\% CL) for a Gaussian CMB map with with the \planck\ characteristics.}
    \label{fig:dust}  
\end{figure}

We used the map of the galactic thermal dust emission at 143 GHz produced by the cleaning method \texttt{Commander} \cite{Eriksen:2007mx} from the \planck\ data and we masked it with the \planck\ 2018 common mask (same as in section \ref{sec:planck}). Even if dust is of course brightest in the galactic plane, the contribution at the non-Gaussian level coming from the rest of the sky is still very large. We computed its integrated bispectrum using again the same 192 step function patches as before ($w_\ell=1$ for $\ell\leq10$ and 0 otherwise) and the results are shown in figure \ref{fig:dust}. We compare the dust integrated bispectrum to the standard deviation expected from a Gaussian CMB map (with the \planck\ characteristics). For every $\ell$, they are of the same order, meaning that the presence of the full dust template in a CMB map could already be seen by eye with the integrated bispectrum (and it was shown in \cite{Jung:2018rgf} that once the information is compressed to the single-parameter $\fnlloc$ the deviation from 0 is more than 5$\sigma$). As in \cite{Jung:2018rgf}, we did not use any linear term to determine this dust template even if the map was masked because the correction is not valid here, the required weak non-Gaussianity approximation is broken here. However, the variance of the dust template seems small enough compared to the signal in figure \ref{fig:dust} (a smooth shape with small fluctations compared to its amplitude), in particular because the CMB variance term in $ C_{\ell_1} C_{\ell_2} C_{\ell_3}$ is not present in this map.

The analysis shown in this section also justifies why we focused on the 143 GHz channel in our previous measurements of the ISW-tSZ-tSZ shape. Even if the dust becomes the dominant galactic foreground at frequencies higher than 100 GHz, its signal is still relatively small at 143 GHz. The dust actually follows a modified blackbody law $I_\nu \propto \nu^{\beta_d} B_\nu(T_d)$ where $B_\nu$ is the Planck's law and the best fit parameters from \planck\ 2018 \cite{Akrami:2018wkt} are $\beta_d=1.48$ (dust spectral index) and $T_d = 19.6$ K (dust temperature). The direct consequence is that the dust bispectrum becomes thousands of times bigger at higher frequencies (like 353 GHz), while the ISW-tSZ-tSZ shape only increases by a factor of a few. This makes very challenging any detection of the latter in the high frequency channels.

\subsection{Forecasts}
\label{sec:forecasts}

In this section, we briefly discuss the future constraints on different shapes, using the integrated bispectrum and focusing on the simple case of an ideal experiment up to $\lmax=3000$.

\begin{figure}
    \centering
    \includegraphics[width=0.66\linewidth]{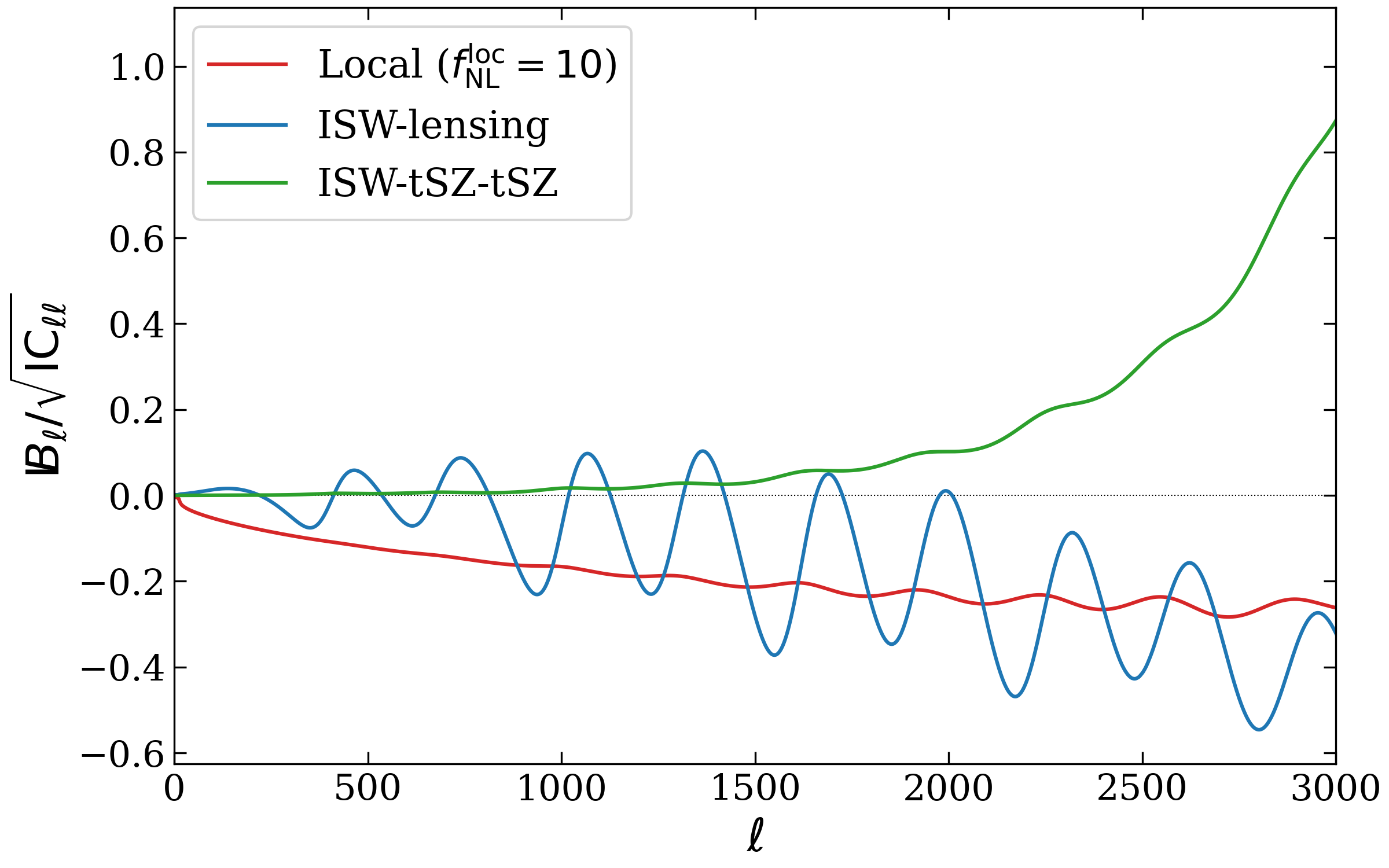}
    \caption{The integrated bispectra divided by the standard deviation (signal-to-noise) of three squeezed bispectrum shapes (local, ISW-lensing, ISW-tSZ-tSZ at 143 GHz). They are computed using eq.\ \eqref{eq:ibisp} (integrated bispectrum) and eq.\ \eqref{eq:ibisp_covariance} (variance) with step function in multipole space patches ($w_\ell = 1$ for $\ell\leq10$, otherwise 0). All the shapes are computed with the \planck\ 2018 best fit cosmology. The local shape is represented in red and the ISW-tSZ-tSZ in green. Note that the local shape is represented for $\fnlloc=10$.}
    \label{fig:Tyy-signaltonoise}  
\end{figure}

In figure \ref{fig:Tyy-signaltonoise}, we show the integrated bispectra of the three squeezed shapes we considered, divided by the standard deviation (signal-to-noise plot). This standard deviation (eq. \ref{eq:ibisp_covariance}), as mentioned at beginning, is computed using \planck\ 2018 cosmological parameters and neglecting instrumental effects (hence it is cosmic-variance limited). The signal-to-noise ratio of the ISW-tSZ-tSZ shape increases strongly above $\ell=2000$, beyond the range of \planck. We note also that we consider a frequency of 143 GHz in this figure. Therefore this ratio can become larger, by taking another frequency channel (2.1 times larger at 100 GHz for example). This confirms the importance of this shape for future experiments. 

\begin{figure}
    \centering
    \includegraphics[width=0.49\linewidth]{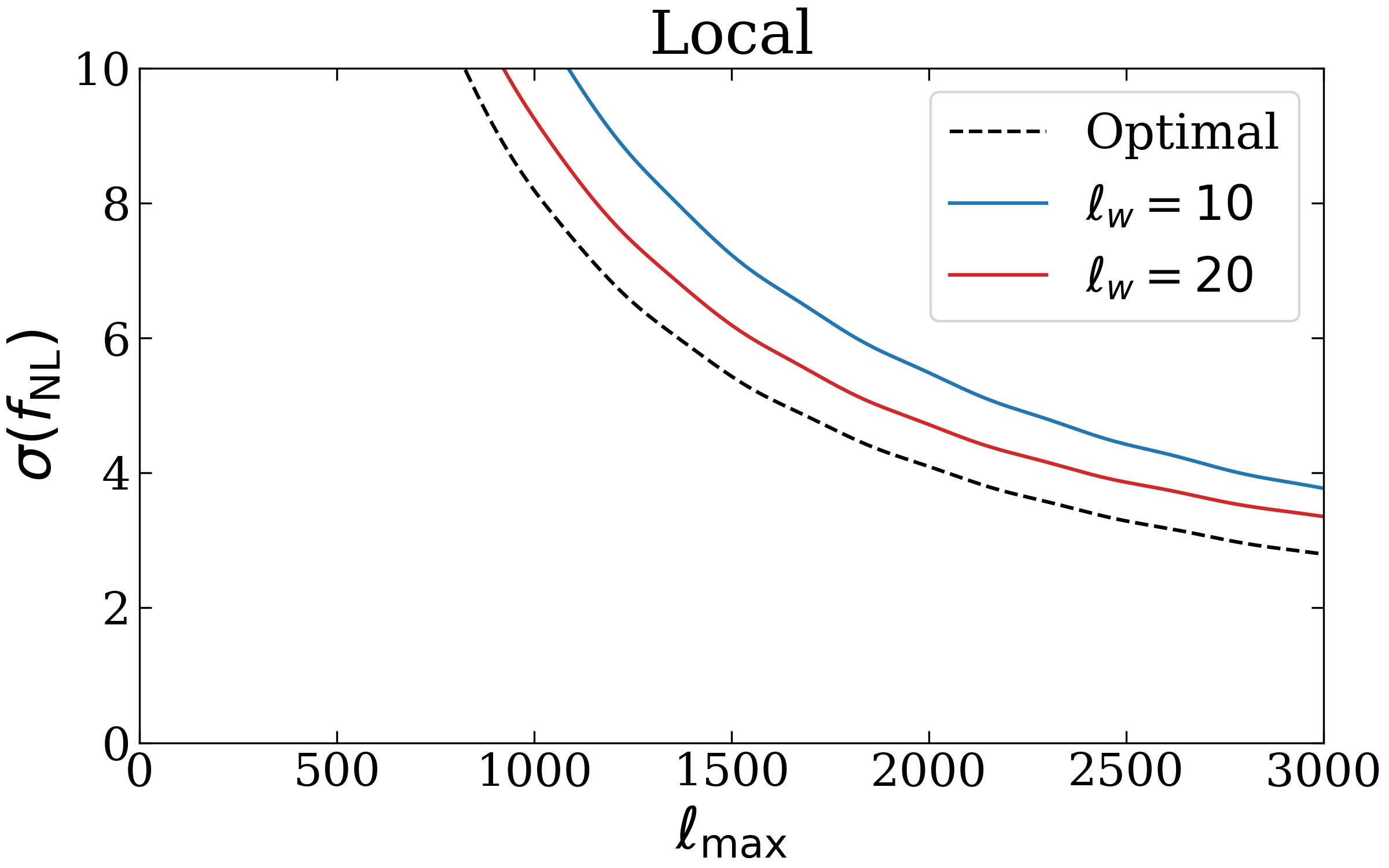}
    \hfill
    \includegraphics[width=0.49\linewidth]{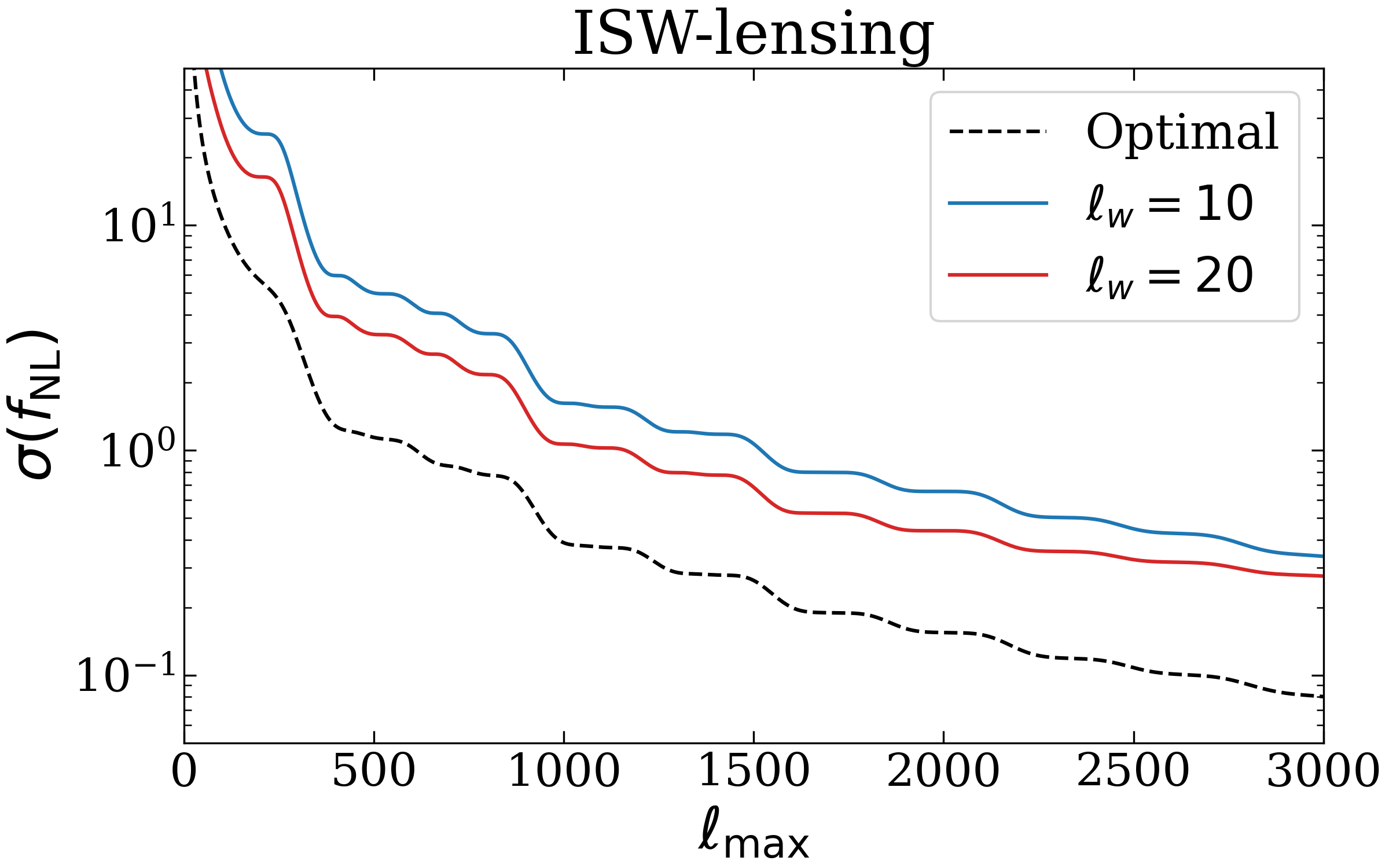}
    \includegraphics[width=0.49\linewidth]{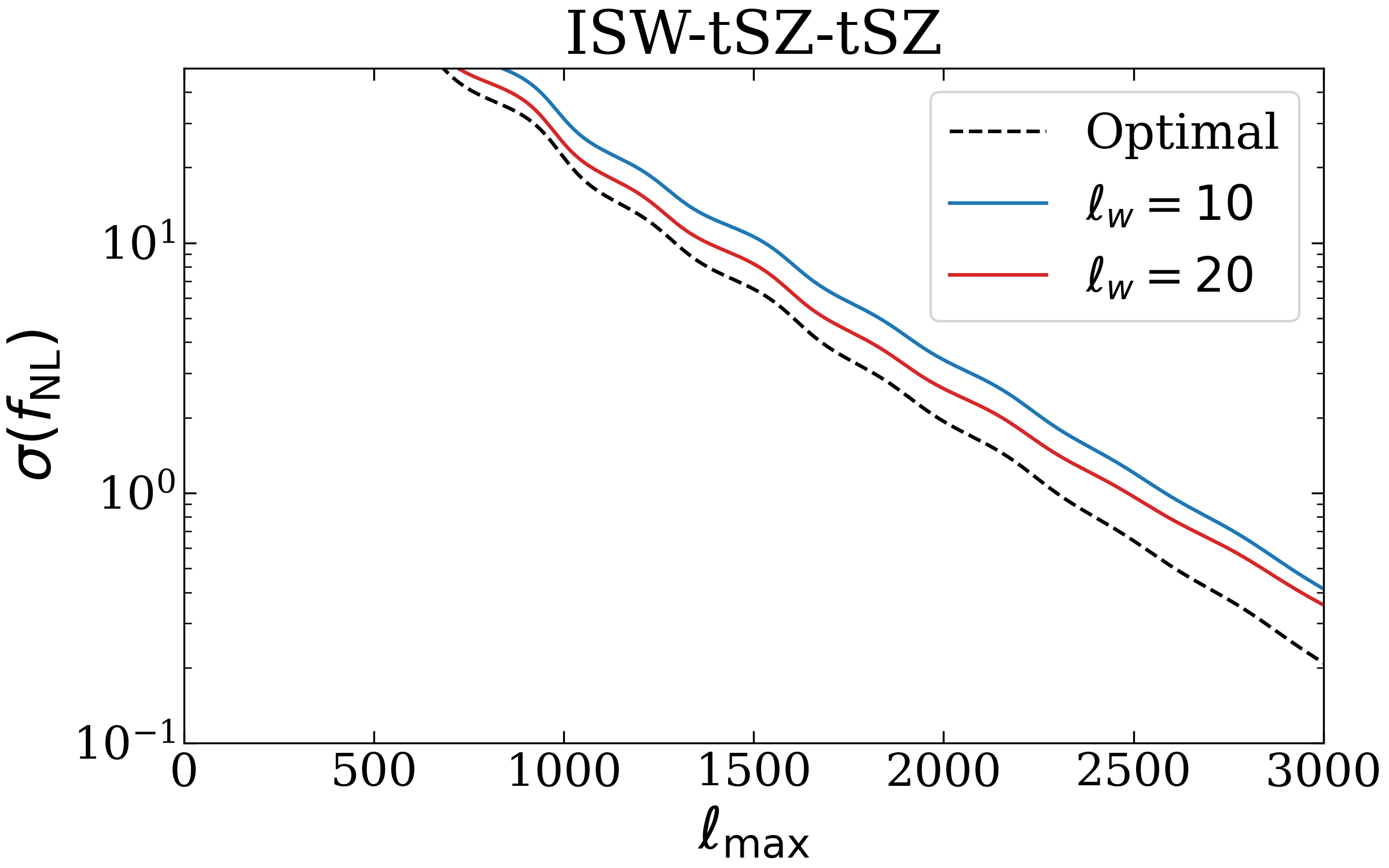}
    \caption{Fisher error bars on $\fnl$ for the local, ISW-lensing and ISW-tSZ-tSZ (143 GHz) shapes in an independent analysis. The blacked dashed lines are the optimal error bars for an estimator using the full bispectrum information. The blue and red solid lines are computed for the integrated bispectrum estimator, using step function patches ($w_\ell = 1$ for $\ell\leq\ell_w$) with $\ell_w=20$ and $\ell_w=10$ respectively. Note the logarithmic scale for the ISW-lensing and ISW-tSZ-tSZ shapes.}
    \label{fig:forecasts}  
\end{figure}

While the error bar on the ISW-tSZ-tSZ template is one order of magnitude too large for a detection in \planck\ data, the large signal-to-noise at high $\ell$ hints for the possibility of a direct detection in future experiments. In figure \ref{fig:forecasts}, we show the Fisher error bars on amplitude parameters for the local, ISW-lensing and ISW-tSZ-tSZ (at 143 GHz) bispectra, both for an optimal bispectrum estimator and the integrated bispectrum estimator, using step function patches ($w_\ell=1$ for $\ell\leq\ell_w$ and 0 otherwise). We also consider two different window function sizes ($\ell_w=10$ and $\ell_w=20$). 

For the local shape, the performance of the integrated bispectrum compared to an optimal estimator stays similar for higher $\lmax$ (roughly 20-25\% suboptimal). Using only CMB temperature data, the error bar on $\fnlloc$ can at most be two times smaller than current constraints, as already pointed out in many bispectrum forecasts. 

A better improvement is possible for the ISW-lensing constraints, but the integrated bispectrum still performs poorly at higher $\lmax$, compared to a full bispectrum analysis. As discussed in section \ref{sec:ISW-lensing}, this is mainly due to the fact that this shape peaks less in the squeezed configuration than the local bispectrum. 

Increasing the size of the window function of the patches ($\ell_w=20$ instead of 10) allows including many more multipole triplets in the integrated bispectrum. However this improves the situation for ISW-lensing only a bit and the results remain very suboptimal. For the local shape, the results can instead improve significantly at $\ell_w=20$, becoming around 10\%, instead of 20\%, suboptimal. However, as detailed in appendix \ref{sec:patch-number}, the practical trade-off is that the number of required patches increases by a factor four when we double $\ell_w$, making the analysis computationally challenging. 

For the ISW-tSZ-tSZ shape, the error bar decreases exponentially at higher resolution. It becomes smaller than $1$ when $\lmax$ is around 2400-2500. As for the local shape, the patches with $\ell_w=20$ only improves a bit the constraints on $\fnl$ and may not be worth the increased computational time, depending on the goal. One can also see that the performance of the integrated bispectrum becomes slightly worse when $\lmax$ increases. At $\lmax=3000$, the integrated bispectrum gives twice as large error bars as the optimal estimator. In table \ref{tab:planck-channels}, they were only 1.5 and 1.7 times bigger in the \planck\ channels 100 and 143 GHz respectively (since the 100 GHz channel has a lower effective $\lmax$ than the 143 GHz one, this is also a confirmation of this trend). For a higher $\lmax$, less squeezed triplets become important also for the ISW-tSZ-tSZ shape. This is also shown in figure \ref{fig:weights}.

\begin{figure}
    \centering
    \includegraphics[width=0.49\linewidth]{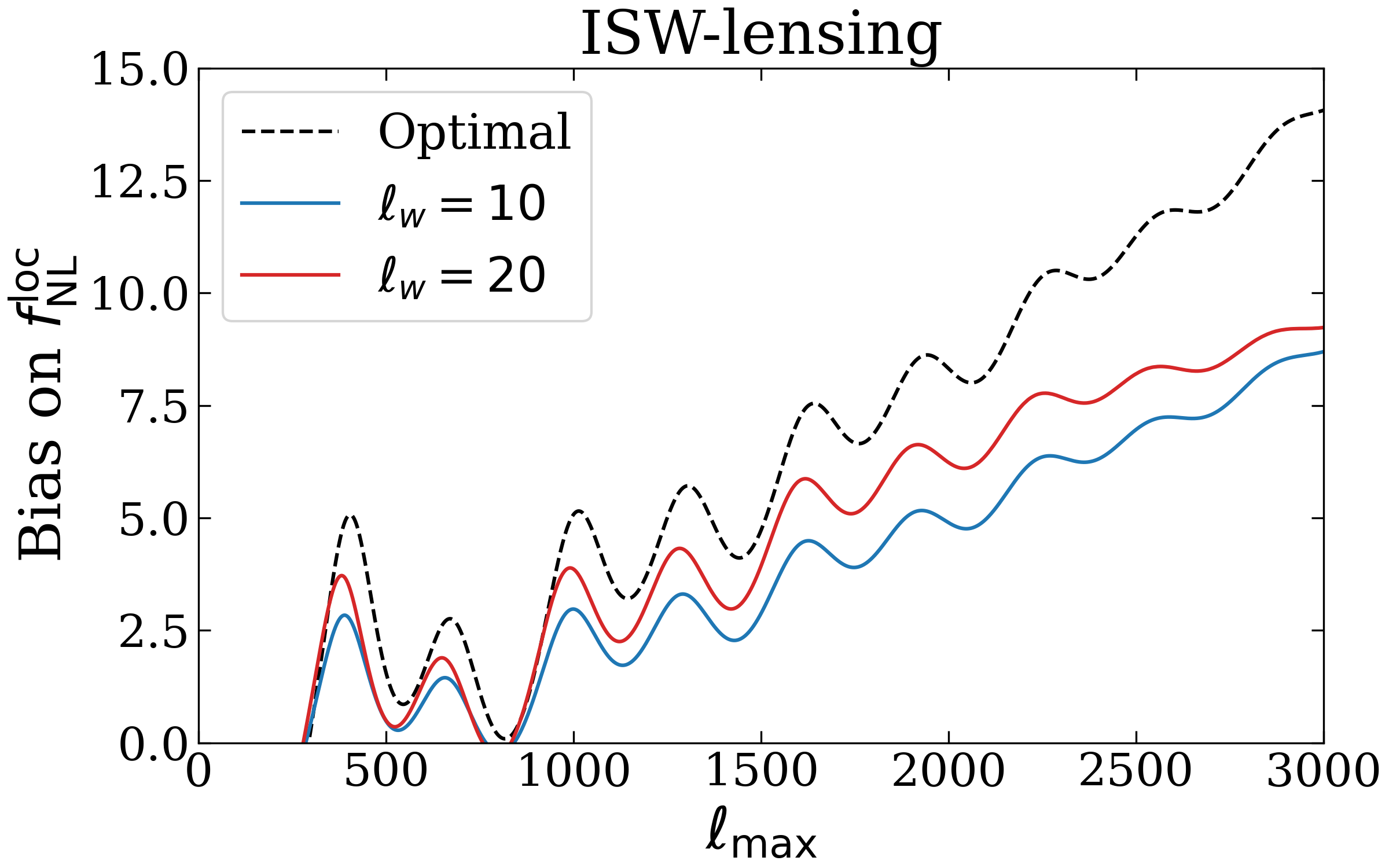}
    \hfill
    \includegraphics[width=0.49\linewidth]{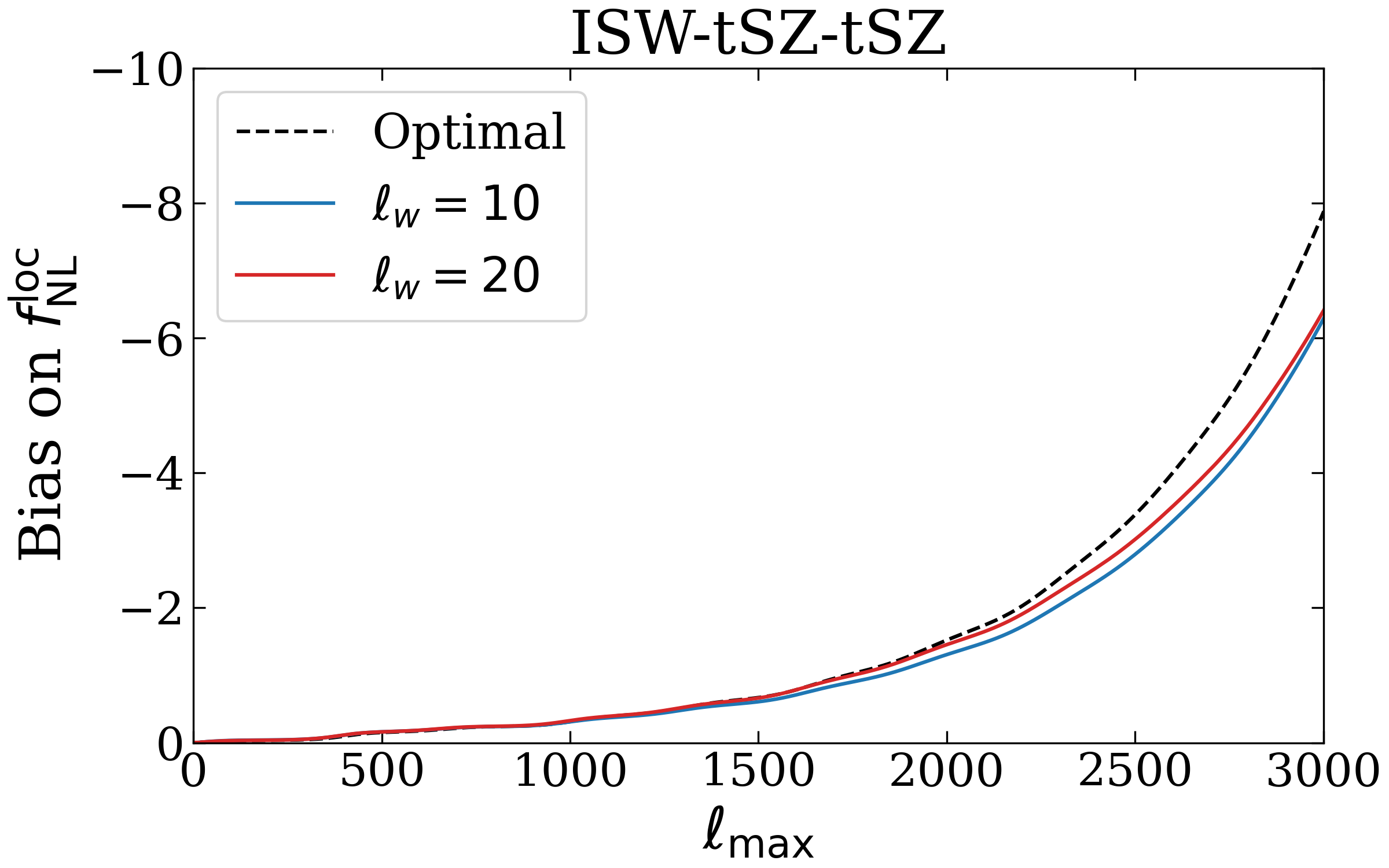}
    \caption{Biases on $\fnlloc$ due to the ISW-lensing (left panel) and ISW-tSZ-tSZ at 143 GHz (right panel) shapes. The blacked dashed lines are computed using the full bispectrum information while the blue and red solid lines are computed for the integrated bispectrum estimator, using step function patches ($w_\ell = 1$ for $\ell\leq\ell_w$) with $\ell_w=20$ and $\ell_w=10$ respectively.}
    \label{fig:bias}  
\end{figure}

When the ISW-tSZ-tSZ shape becomes detectable (large $\lmax$), the bias it produces on the local shape also increases. The two shapes are very correlated (because they are both very squeezed), therefore ISW-tSZ-tSZ produces a deviation at the level of  several $\sigma(\fnlloc)$ at $\lmax=3000$, as shown in figure \ref{fig:bias} and as pointed out originally in \cite{Hill:2018ypf}. For a low $\lmax$, the integrated bispectrum and the full bispectrum biases are very close. The difference only appears at higher $\lmax$. This is again in agreement with the discussion on error bars above. The integrated bispectrum efficiency does not change with $\lmax$ for the local shape, but decreases for the ISW-tSZ-tSZ template meaning the integrated bispectrum will start to pick less information for this shape than for the local template when $\lmax$ increases and thus the bias becomes smaller. The bias produced by the ISW-lensing correlations is always smaller in the integrated bispectrum case than for the bispectrum for the same reason (for every $\lmax$, the integrated bispectrum works better with the local shape than with the ISW-lensing template).

In both figures \ref{fig:forecasts} and \ref{fig:bias}, the ISW-tSZ-tSZ shape is computed in the single-frequency channel 143 GHz. The corresponding bias and error bar should respectively be multiplied or divided by the factor $g(\nu)^2/g(143~\mathrm{GHz})^2$ to be converted to another frequency channel. Hence they become respectively larger and smaller for a smaller frequency channel (2.1 times at 100 GHz), or for a sufficiently higher frequency (4.6 at 353 GHz) where other issues like the dust contamination makes impossible to reach optimality with a large $\lmax$.

\section{Conclusions}
\label{sec:conclusion}

In this work we study the integrated bispectrum method for 2D cosmological fields on the sphere, without resorting to any flat sky approximation.
The integrated angular bispectrum, in analogy to its 3D counterpart, provides compressed information about squeezed angular bispectrum configurations, by measuring the large scale modulation of the angular power spectrum from separate, small patches of the sky.

In the first part of the paper we derive useful analytical formulae for the full sky integrated angular bispectrum ($\ib_{\ell}$) and its covariance (eqs.\ \ref{eq:ibisp} and \ref{eq:ibisp_covariance}, respectively), in which the $m$-dependence is eliminated by choosing suitable, azimuthally symmetric patches. This allows for fast computation of the expected slope of $\ib_{\ell}$ and makes both data analysis and comparisons between actual measurements and theoretical expectations much easier.
We then consider the construction of integrated angular bispectrum estimators, in the realistic situation in which isotropy is broken by effects such as non-stationary noise, or the presence of a sky cut. As it is well known, isotropy breaking produces a spurious increase in the error bars of the estimator, with respect to the full sky, isotropic case. Such increase can be removed by introducing a suitable mean-field correction term. We explicitly build this correction (eq.\ \ref{eq:linear-correction}), in analogy to what done in the literature for general CMB bispectrum and lensing estimators.

In the second part of the paper, we implement the methodological findings summarized above to build a pipeline for integrated angular bispectrum estimation, forecasting and fitting, and we show some applications of this pipeline, by focusing specifically on CMB data analysis. In one of such applications, we estimate the integrated angular bispectrum from \planck\ data and fit it to theoretical $\ib_{\ell}$ predictions, in presence of local primordial NG. This allows us to compare and validate our results against those obtained by the \planck\ team by mean of optimal bispectrum template-fitting methods. At the same time, it provides an additional measurement of $\fnlloc$, which is slightly sub-optimal but has a very transparent physical interpretation (i.e.\ we measure directly the real space, large scale modulation of small scale power, which is expected in presence of squeezed NG). In similar fashion, we produce an $\ib_{\ell}$ analysis of the ISW-lensing bispectrum, which was actually detected by \planck\ . We find in this case that the increase in error bars, compared to optimal estimates of the ISW-lensing amplitude, is much larger than for the local case (the shape takes non-negligible contributions from non-squeezed triplets). As a result, our $\ib_{\ell}$ estimator is unable to detect a significant signal, contrary to optimal bispectrum estimators. Both our local and ISW-lensing analyses are validated via extensive tests on simulations.

We then consider a third shape, namely the bispectrum arising from the coupling between ISW and the thermal Sunyaev Zeldovich effect (ISW-tSZ-tSZ). This was recently suggested as a potential contaminant for primordial local NG measurements, which deserves specific attention \cite{Hill:2018ypf}. We study this shape in detail, using a halo model approach similar to that of \cite{Hill:2018ypf}, and we derive the expected $\ib_{\ell}$ arising from it. We obtain that the expected amplitude of this bispectrum signal is too small to make it a significant contaminant of local $\fnl$ measurements, at \planck\ resolution and noise level. 

We also build actual estimators of the ISW-tSZ-tSZ amplitude (i.e., we focus on ISW-tSZ-tSZ as an interesting NG signal per se, rather than as a spurious primordial $\fnl$ contaminant, via its correlation with the local shape). Since the ISW-tSZ-tSZ bispectrum strongly peaks in the squeezed limit, we find that an $\ib_{\ell}$ template-fitting estimator is in this specific case not only easier to implement, but also nearly as optimal as a full bispectrum estimator. Confirming theoretical predictions, we are however unable to detect the ISW-tSZ-tSZ signal using \planck\ $100$ and $143$ GHz channels, due to its small amplitude. We produce simple forecasts, where we consider a single, noiseless frequency channel and neglect issues arising from foreground contamination. On the basis of such forecasts, we expect that a detection of ISW-tSZ-tSZ will remain quite challenging, but not fully impossible, for future surveys.

Finally, we argue that the integrated angular bispectrum can be used as a fast and simple to estimate, yet powerful observable to diagnose spurious foreground contamination in CMB data, via NG signatures. We show a preliminary example by extracting the $\ib_{\ell}$ of the dust \texttt{Commander} map.

We conclude by stressing again that, while in this work we discussed only CMB data analysis examples, our pipeline can be applied to any cosmological 2D-field on the sphere, such as e.g. weak lensing convergence or projected galaxy maps, considering both auto and cross-spectra. This will be the object of future projects.
\\

{\bf Acknowledgements:} Some of the results in this paper have been derived using the healpy \cite{Zonca2019} and HEALPix packages. We also acknowledge the use of CAMB. 

\noindent The authors thank Boris Bolliet for useful discussions about the tSZ effect. The authors also thank Bartjan van Tent for valuable discussions on bispectrum estimation.

\noindent GJ's, FO's and ML's work was supported by the University of Padova under the STARS Grants programme CoGITO, Cosmology beyond Gaussianity, Inference, Theory and Observations. GJ, FO and ML also acknowledge support from the ASI-COSMOS Network (cosmosnet.it) and from the INDARK INFN Initiative (web.infn.it/CSN4/IS/Linea5/InDark), which provided access to CINECA supercomputing facilities (cineca.it)

\noindent GJ and ML were supported by the project "Combining Cosmic Microwave Background and Large Scale Structure data: an Integrated Approach for Addressing Fundamental Questions in Cosmology", funded by the MIUR Progetti di Ricerca di Rilevante Interesse Nazionale (PRIN) Bando 2017 - grant 2017YJYZAH.

\noindent AR was supported by the ERC Consolidator Grant {\it CMBSPEC} (No.~725456) as part of the European Union's Horizon 2020 research and innovation program.

\appendix
\section{Choice of patches}

In this appendix we discuss the choice of patches we used for the different analyses presented in this paper: a set of 192 step function in multipole space patches ($w_\ell=1$ for $\ell\leq\ell_w$, with $\ell_w=10$, see eq.\ \ref{eq:patch-harmonic}).

\subsection{Optimal number of patches}
\label{sec:patch-number}

Each multipole space patch, as introduced in section \ref{sec:multipole-patch}, is defined by its central position $\hat{\Omega}_0$ and the corresponding window function in harmonic space (eq.\ \ref{eq:patch-harmonic}). Such a patch is the largest around its center in a region where resides most of its constraining power, but it is also non-zero in the rest of the sky. Then, one needs to use enough of these patches, with a uniform distribution over the sky of their central positions, to obtain an isotropic coverage of the sky because it is important not to introduce any additional anisotropy. Unlike the HEALPix patches, which we will discuss in appendix \ref{sec:healpix-patch}, where the sky is divided into a certain number of equal-sized areas and thus directly giving the number of required patches, this number is not so straightforward to get for multipole space patches. This section is devoted to the tests we performed to verify that our choice was the most optimal one for our cases of study.

We recall that the method we used for the uniform repartition of their centers was simply to place eeach of them at the center of a corresponding HEALPix patch. This implies that we have to use a number of patch $N_\mathrm{patches}$ corresponding to the number of pixels of a low resolution map ($N_\mathrm{patches}=48$, 192 or 768). It is important to keep $N_\mathrm{patches}$ as small as possible because it is proportional to the amount of computations (one power spectrum to determine per patch) while of course still getting optimal results in agreement with the theory. The effect of not using enough patches is roughly similar to a partial sky coverage in the sense it will also increase the variance because some parts of the sky are less scanned than others.

\begin{figure}
    \centering
    \includegraphics[width=\textwidth]{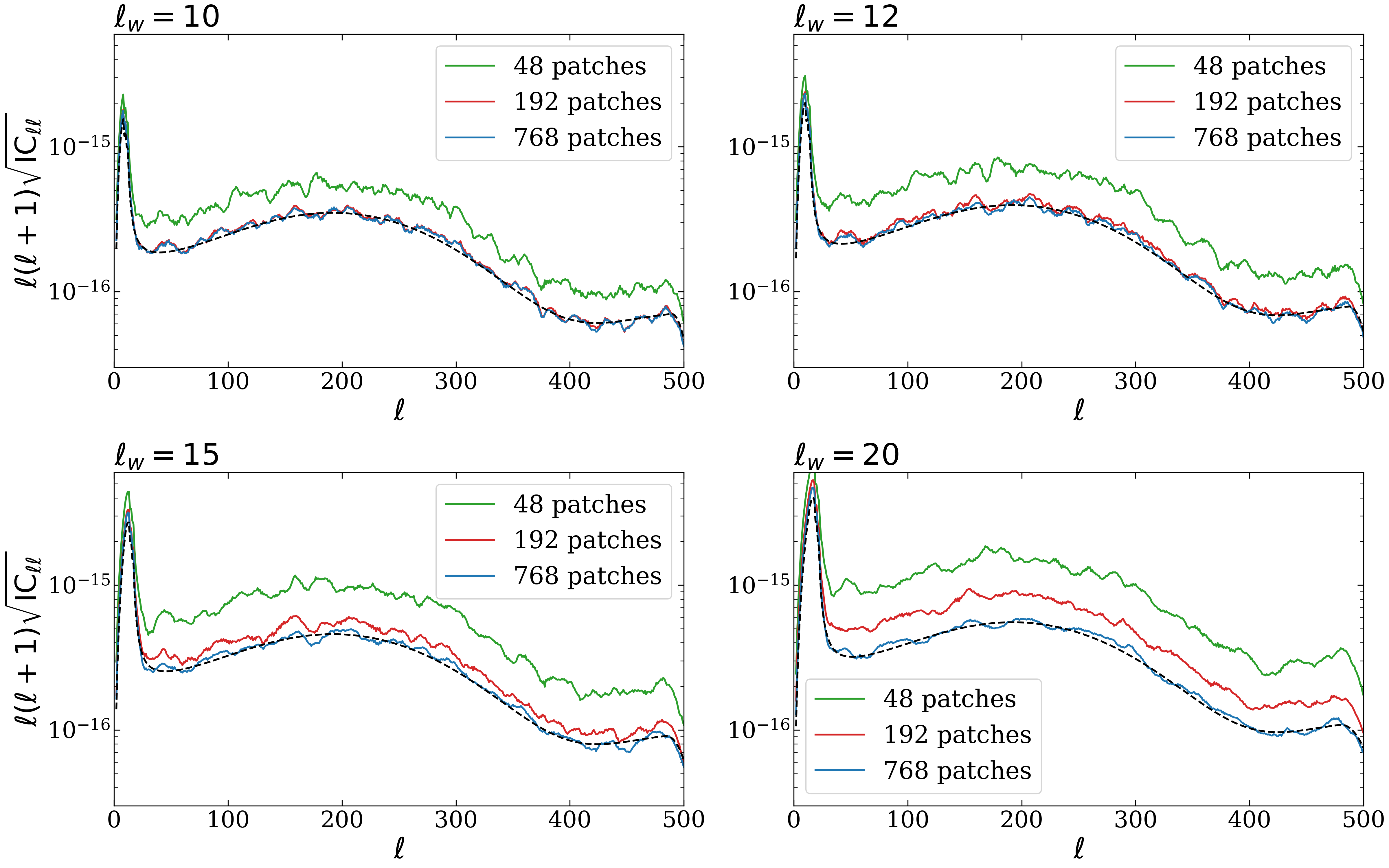}
    \caption{The standard deviation of the integrated bispectrum from 100 Gaussian CMB maps determined using step function patches ($w_\ell=1$ for $\ell\leq\ell_w$) for different window function sizes ($\ell_w=10$, 12, 15 and 20). The black dashed lines are the theoretical expectations computed using eq.\ \eqref{eq:ibisp_covariance} while the solid green, red and blue lines correspond respectively to a number of patches of 48, 192 and 768. Note the logarithmic scale on the vertical axis.}
    \label{fig:windows}  
\end{figure}

Our simple test consisted on comparing the standard deviation of the integrated bispectrum determined from a set of 100 Gaussian CMB realizations to its theoretical counterpart. In figure \ref{fig:windows}, we show the results for three different number of patches ($N_\mathrm{patches}=48$, 192 and 768) and four different window function sizes ($\ell_w=10$, 12, 15 and 20). Optimality is never reached with 48 patches. For $\ell_w=10$, we obtained optimal results with both 192 and 768 patches, justifying our choice of the smaller value for the analyses performed in this paper. For larger $\ell_w$, we can see the increase in variance with 192 patches, meaning that it is required to use 768 patches above $\ell_w=10$. For $\ell_w=20$, the results with 768 patches are still optimal (for even larger $\ell_w$, which are not shown here, 768 patches are also not sufficient). Obviously, a larger window function will include more bispectrum configurations and leads to better constraints on $\fnl$. Depending on the studied shapes, it can be interesting to use the larger window function, which however increases the computational time by a factor four. Intermediate cases like $\ell_w=15$ are a priori not useful because they require as many calculations as $\ell_w=20$, for weaker constraints. 

\subsection{HEALPix patches}
\label{sec:healpix-patch}

Other types of patches than the step function ones can also be chosen to perform similar analyses. Here we discuss the simple case of HEALPix patches, introduced in section \ref{sec:implementation} and shown in the top panels of figure \ref{fig:patches}.

\begin{figure}
    \centering
    \includegraphics[width=0.66\textwidth]{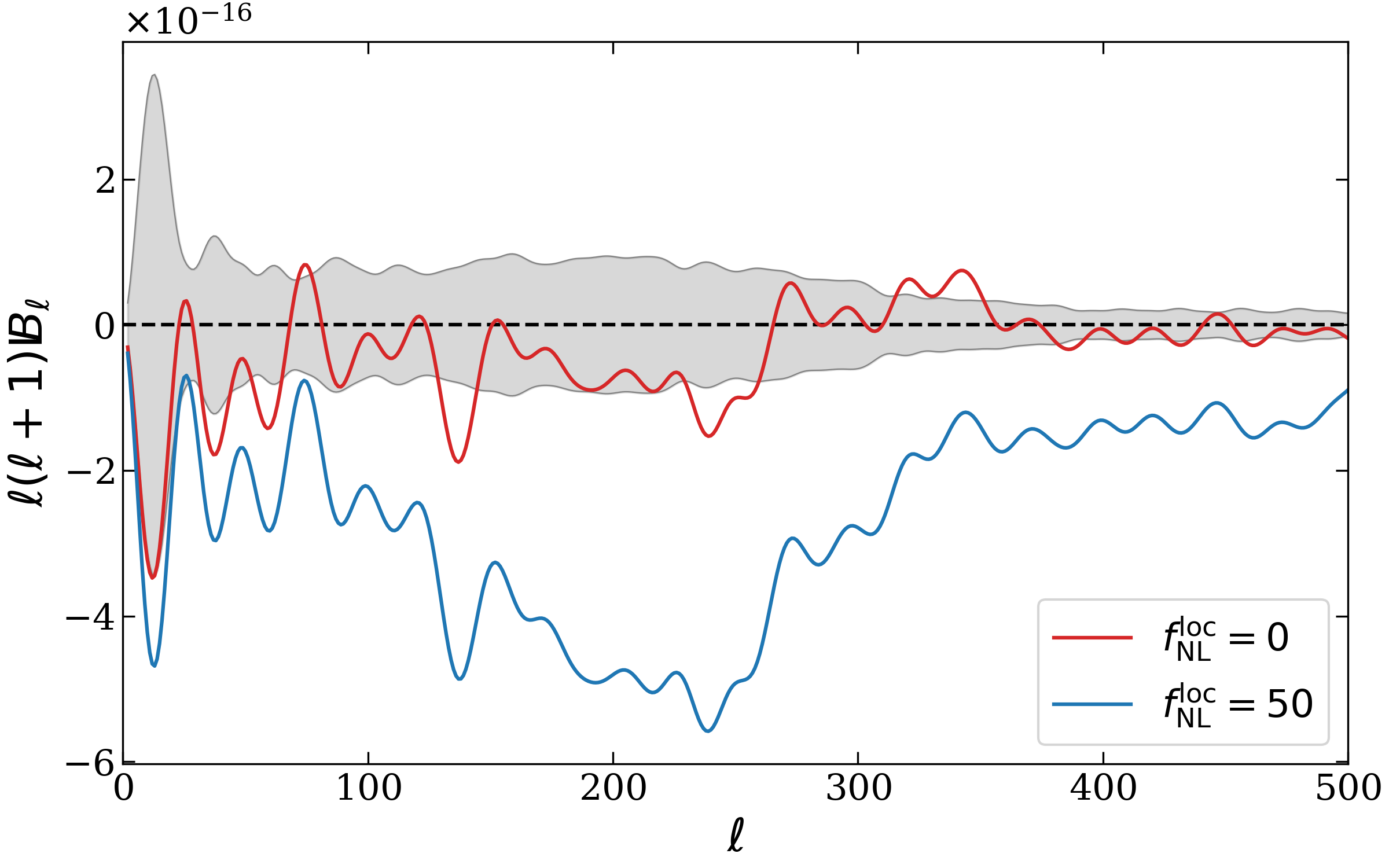}
    \caption{The integrated bispectra from CMB simulations ($N_\mathrm{side}=256$) using 192 HEALPix patches. The averaged integrated bispectra from 100 Gaussian maps $\fnlloc=0$ in red and from 100 non-Gaussian maps $\fnlloc=50$ in blue. The standard error around zero in grey is also determined from the data at there is no theoretical prediction in that case.}
    \label{fig:loc-256-healpix}  
\end{figure}

For this analysis, the sky is divided into 192 equal parts built using the pixelization technique of HEALPix. We study the same 100 Gaussian CMB maps as in the previous section and also 100 non-Gaussian maps (having the same linear part) with $\fnlloc=50$. Except for the lower number of maps considered, this is similar to the first test described in section \ref{sec:local}. Figure \ref{fig:loc-256-healpix} shows the averaged integrated bispectra determined from these two sets. However, with HEALPix patches we cannot compute theoretical shapes and the expected covariance. They could however be inferred from a large number of simulations. For example, we indicated the standard deviation of the integrated bispectrum determined from the 100 Gaussian maps. It is clear that 100 maps are not sufficient to correctly estimate the shape and a larger number of maps should be used, which may be doable at low resolution like here, but a lot more difficult at $\nside=2048$. For the non-Gaussian maps, we can see the first acoustic peak of the power spectrum as with the step function patches.

\subsection{Mexican needlet patches}
\label{sec:mexican-patch}

For a better localization in real space of multipole space patches, one can use mexican needlets defined by
\begin{equation}
    \label{eq:mexican-needlets}
    w_\ell = \left[\frac{\ell(\ell+1)}{B^{2j}}\right]^p e^{\frac{-\ell(\ell+1)}{B^{2j}}}\,,
\end{equation}
where $B$, $j$ and $p$ are the needlet parameters determining its properties (for a complete discussion on the topic, see \cite{2011ApJ...733..121S}).

We used the following parameters $p=1$, $B=4$ and $j=1$ for which $w_\ell$ peaks below 10 and is effectively 0 at $\ell_w=10$, allowing us to consider it is zero for larger $\ell$'s. Figure \ref{fig:patches} shows three such patches on the bottom line.

\begin{figure}
    \centering
    \includegraphics[width=\textwidth]{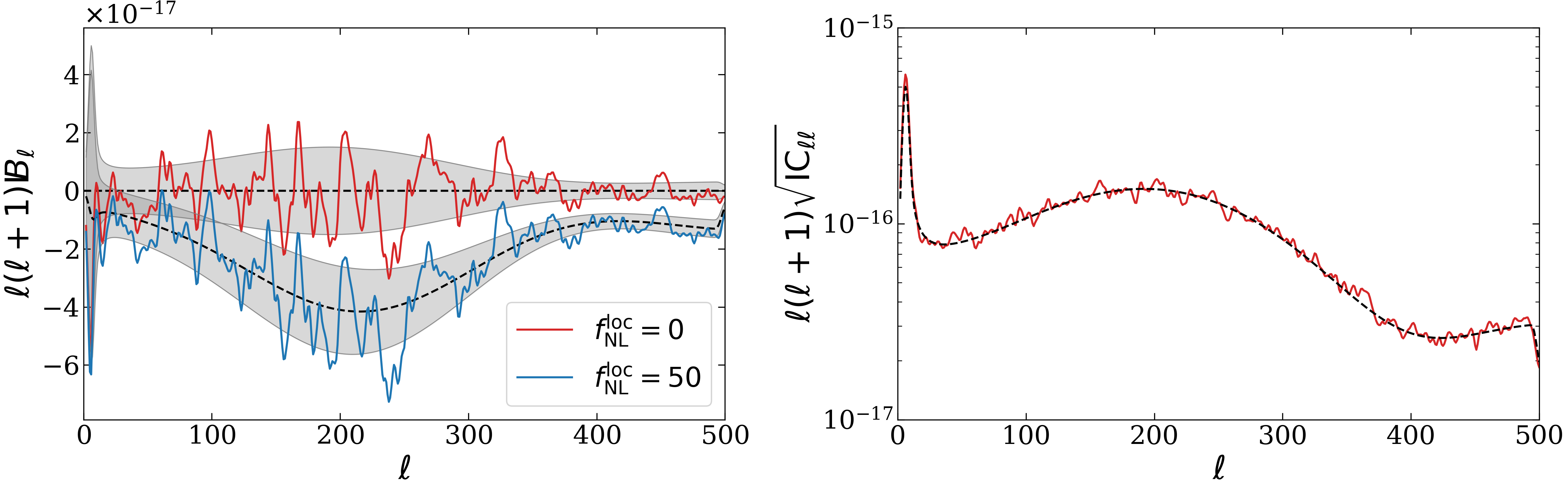}
    \caption{The integrated bispectra from CMB simulations ($\nside=256$) using 192 mexican needlet patches ($\ell_w=10$, $p=1$, $B=4$ and $j=1$). On the left, the averaged integrated bispectra from 100 Gaussian maps $\fnlloc=0$ in red and from 100 non-Gaussian maps $\fnlloc=50$ in blue. The black dashed lines correspond to their respective theoretical predictions computed using eq.\ \eqref{eq:ibisp} with their expected standard error (68\% CL) in grey, computed with eq.\ \eqref{eq:ibisp_covariance}. On the right, the standard deviation from the 100 Gaussian maps in red and its theoretical counterpart in black (dashed line). Note that all quantities are multiplied by the factor $\ell(\ell+1)$ for visibility, and that the scale is logarithmic in the right panel.}
    \label{fig:loc-256-mex}  
\end{figure}

Applying again the pipeline described in section \ref{sec:implementation}, we obtained from the data the averaged integrated bispectra shown in figure \ref{fig:loc-256-mex}. They fit perfectly the theoretical predictions. The integrated bispectrum estimator (eq.\ \ref{eq:fnl}) can also be used on these maps, and doing so we obtained $\fnlloc = 0.6 \pm 2.6$ for the Gaussian maps and $\fnlloc = 52.9 \pm 2.9$ for the non-Gaussian maps, in agreement with the expected results. The error bars indicated here are the standard errors determined from the 100 maps. Hence, the standard deviations on one map for $\fnl$ are ten times larger (26 and 29). These values are larger than the ones found with the step function patches (21 against 26 for the Gaussian maps), thus the mexican needlet patches (with these parameters) are suboptimal compared to the step function patches we used in section \ref{sec:data}.

This concludes the section, showing that other choices of patches can be considered as the method we developed is more general than the examples detailed in the paper.

\section{Bispectrum weights in multipole space}
\label{sec:weights}

We discuss here the alternative representation of the main squeezed bispectrum shapes (i.e.\ local, ISW-lensing, ISW-tSZ-tSZ) as shown in figure \ref{fig:weights}, where the information is less-compressed than with the integrated bispectrum.

We recall that the optimal estimator of $\fnl$ is given by
\begin{equation}
    \label{eq:bispectrum-estimator}
    \hat{f}_\mathrm{NL} = \frac{\sum\limits_{\ell_1\leq\ell_2\leq\ell_3} 
        \frac{B_{\ell_1\ell_2\ell_3}^\mathrm{obs} B_{\ell_1\ell_2\ell_3}^\mathrm{th}}{V_{\ell_1\ell_2\ell_3}}} {\sum\limits_{\ell_1\leq\ell_2\leq\ell_3} \frac{(B_{\ell_1\ell_2\ell_3}^\mathrm{th})^2 }{V_{\ell_1\ell_2\ell_3}}}
        = \sum\limits_{\ell_1\leq\ell_2\leq\ell_3} w_{\ell_1\ell_2\ell_3} \frac{B_{\ell_1\ell_2\ell_3}^\mathrm{obs}}{B_{\ell_1\ell_2\ell_3}^\mathrm{th}},\quad
        \text{with}~~
    w_{\ell_1\ell_2\ell_3} = \frac{\frac{(B_{\ell_1\ell_2\ell_3}^\mathrm{th})^2}{V_{\ell_1\ell_2\ell_3}}}
        {\sum\limits_{\ell_1\leq\ell_2\leq\ell_3} \frac{(B_{\ell_1\ell_2\ell_3}^\mathrm{th})^2}{V_{\ell_1\ell_2\ell_3}}}\,,
\end{equation}
where $V_{\ell_1\ell_2\ell_3}$ is the bispectrum variance given in eq.\ \eqref{eq:bispectrum-covariance}. The first part of this equation is the classic result of \cite{Komatsu:2001rj} while the second part is an alternative form illustrating the fact that this estimator is simply a weighted summation of the $\fnl$-estimator for a single multipole triplet $B_{\ell_1\ell_2\ell_3}^\mathrm{obs} / B_{\ell_1\ell_2\ell_3}^\mathrm{th}$ (see \cite{Bucher:2009nm}). The weights $w_{\ell_1\ell_2\ell_3}$ are thus useful to visualize where a bispectrum shape peak in multipole space and to determine the multipole triplets which matter the most to estimate $\fnl$.

\bibliographystyle{JHEP}
\bibliography{biblio}

\end{document}